\documentclass[12pt, aip,showpacs,superscriptaddress]{revtex4-1}
\usepackage{fourier}
\usepackage{url,ulem}
\usepackage{amsmath,amssymb}
\usepackage{graphicx,color}

\begin{document}

\title{Full particle orbit effects in regular and stochastic magnetic fields}

\author{Shun Ogawa}
\email{shun.ogawa@cpt.univ-mrs.fr}
\affiliation{Aix Marseille Univ., Univ. Toulon, CNRS, CPT, Marseille, France}
\affiliation{CEA, IRFM, F-13108 St. Paul-lez-Durance cedex, France }

\author{Benjamin Cambon}
\affiliation{Aix Marseille Univ., Univ. Toulon, CNRS, CPT, Marseille, France}

\author{Xavier Leoncini}
\affiliation{Aix Marseille Univ., Univ. Toulon, CNRS, CPT, Marseille, France}

\author{Michel Vittot}
\affiliation{Aix Marseille Univ., Univ. Toulon, CNRS, CPT, Marseille, France}

\author{Diego del Castillo-Negrete}
\affiliation{Oak Ridge National Laboratory, Oak Ridge, TN, USA}

\author{Guilhem Dif-Pradalier}
\affiliation{CEA, IRFM, F-13108 St. Paul-lez-Durance cedex, France }

\author{Xavier Garbet}
\affiliation{CEA, IRFM, F-13108 St. Paul-lez-Durance cedex, France }


\begin{abstract}
We present a numerical study of charged particle motion in a time-independent magnetic field in cylindrical geometry. 
The magnetic field model consists of an unperturbed reversed-shear (non-monotonic $q$-profile) helical part  and a perturbation consisting of a superposition of modes.
 Contrary to most of the previous studies, the particle trajectories  are computed by directly solving the full Lorentz force equations of motion in a six-dimensional phase space using a sixth-order, implicit, symplectic Gauss-Legendre method. 
The level of stochasticity in the particle orbits is diagnosed using averaged, effective Poincare sections. 
It is shown that when only one mode is present the particle orbits can be stochastic even though the magnetic field line orbits are not stochastic (i.e. fully integrable). 
The lack of integrability of the particle orbits in this case is related to separatrix crossing and the breakdown of the global conservation of the magnetic moment. 
Some perturbation consisting of two modes creates resonance overlapping, leading to Hamiltonian chaos
in magnetic field lines. 
Then, the particle orbits exhibit a nontrivial dynamics depending on their energy and pitch angle. 
It is shown that the regions where the particle motion is stochastic decrease as the energy increases. 
The non-monotonicity of the $q$-profile implies the existence of magnetic ITBs 
(internal transport barriers) which correspond to shearless flux surfaces located in the vicinity of the $q$-profile minimum. 
It is shown that depending on the energy, these magnetic ITBs might or might not confine particles. 
That is, magnetic ITBs  act as an energy-dependent particle confinement filter. 
Magnetic field lines in reversed-shear configurations exhibit topological bifurcations (from homoclinic to heteroclinic) due to separatrix reconnection. 
We show that a similar but more complex scenario appears in the case of particle orbits that depends in a non-trivial way on the energy and pitch angle of the particles. 
 \end{abstract}
\maketitle

\section{Introduction}

The motion of a charged particle evolving in a complex magnetic field
is investigated with magnetized plasma confinement in mind. 
This problem is a long standing issue with many studies, most of them relying on
guiding center and gyrokinetic reductions \cite{Boozer2004,CaryBrizard2009,Littlejohn1981}.
These reductions typically rely on a separation of scales both in space and time. 
For instance, one assumes that a spatial scale of the cyclotron motion, 
the Larmor radius $r_{{\rm L}}$, is small enough compared with the characteristic
scale $L$ of the electromagnetic field, and that the cyclotron frequency
is much higher than any other characteristic frequency in the plasma \cite{Boozer2004, CaryBrizard2009}. 
In the guiding center theory, the degrees of freedom
in a configurational space are reduced by averaging over a gyroperiod,
the degree of freedom corresponding to the gyrophase is wiped out.
The conjugate of the gyrophase is the magnetic moment, and it becomes
an adiabatic invariant. In the gyrokinetic approach, gyrokinetic coordinates
are introduced starting from the guiding center coordinates as a first
order approximation, eventually through for instance the use of a
Lie-transform method\cite{Lichtenberg}, it as well assumes that the newly obtained
magnetic moment $\mu_{\rm gyro}$ becomes an invariant\cite{CaryBrizard2009}. 
Thus, the dimension of the original system is reduced and in principle 
the integration gets to be easier, most notably the computational effort
to perform realistic kinetic simulations of the plasma is greatly reduced, 
and becomes accessible to current accessible computing facilities. 
This explains why many studies on magnetized plasmas 
are based on numerical
codes that integrate up to some accuracies the gyrokinetic equations.

On the other hand, the increased computing power allows us as well
to integrate and investigate long time particle trajectories 
with small time slice compared with the cyclotron motion, 
and reach for instance the adiabatic time scales, and this is the approach
we consider in this paper. 
To be more specific, we shall consider
the motion of a charged particle in a fixed external magnetic field,
and deal with the case in which the assumption of the guiding center
theory breaks down.
Similar phenomena have been discussed by 
Landsman \textit{et al.} \cite{Landsman2004}, 
and more recently by Cambon \textit{et al.} \cite{Cambon2014}.

In this setting we can exhibit some potentially
physical relevant effect beyond gyrokinetic or guiding center reductions. 
For instance, Recently Pfefferl\'e {\it et al.}\cite{Pfefferle2015} have found that 
there exists a case where the full trajectory and 
the trajectory obtained from guiding center reduction are completely 
different from each other even though the gradient of the magnetic field is $0$.
In the same vein, Cambon \textit{et al.} \cite{Cambon2014} study the full orbit effects of charged particle
motion in a magnetic field in toroidal and cylindrical geometry without
any assumptions of invariance of the magnetic moment. 
Signature of Hamiltonian chaos in phase space 
was clearly exhibited, even when the magnetic field is toroidally symmetric.
This effect is enhanced when a non-generic ripple effect that has no radial element and
keeps field lines integrable is introduced. 
In the chaotic regions, the magnetic moment $\mu$
is no longer a constant of motion, and the basic assumptions of the
gyrokinetic theory or guiding center theory breaks down. 
In order to observe the chaotic motion, a naive Poincar\'e plot based on spatial periodicity is unconvincing. 
Indeed, the Larmor gyration blurs the
Poincar\'e plot, and we cannot really conclude on whether or not the
motion is chaotic.
In Cambon {\it et al.}\cite{Cambon2014}, Poincar\'e plots were taken
for each time when $\mu=\langle\mu\rangle_{{\rm time}}$ or when $d\mu/dt=0$
and $d^{2}\mu/dt^{2}>0$, where $\langle\bullet\rangle_{{\rm time}}$
denotes the time average. Further, they also considered that the Hamiltonian
$H_{R_{{\rm tor}}}$ of the toroidal system with radius $R_{{\rm tor}}$
consists of the cylinder part $H_{\infty}$ corresponding to the infinite
toroidal radius and the perturbation part $H_{{\rm curv}}\sim1/R_{{\rm tor}}$
corresponding to a finite curvature. The Hamiltonian $H_{\infty}$
is conserved when $R_{{\rm tor}}=\infty$, and oscillates when $R_{{\rm tor}}<\infty$.
Then Poincar\'e plots are taken on the iso-$H_{\infty}$ set, and the
chaos induced with the toroidal geometry are shown. 
In this paper, we consider at first the same problem but in
a different geometry, namely a purely cylindrical one. 
The onset of chaos being triggered by introducing gradually a perturbation in the magnetic field. 
This allows us to investigate even more the complex relationship 
between the chaos of field lines and particle motion.
As a consequence of this we examine the feasibility of setting upon 
internal transport barrier (ITB)\cite{Connor2004, Constantinescu2012}   
using given magnetic configurations.

To be more specific, 
of particular interest to the present paper is the study of the role of finite Larmor radius effects in reversed shear magnetic fields, 
that is, magnetic configurations with non-monotonic $q$-profiles known to exhibit ITBs.  
The formation of ITB is a complex, not fully understood process involving several physics mechanics. 
However,  at the heart of this problem there is a fundamental Hamiltonian dynamical systems problem related 
to the perturbation of degenerate Hamiltonian systems with the Hamiltonian $H_0$. 
In the standard (non-degenerate) case, integrable Hamiltonians exhibit a monotonic dependence on the action. 
In the one-dimensional case, this implies that the derivative of the unperturbed frequency with respect to the action,
$d^2H_0/d J^2$, never vanishes.  
In $N$-dimensions this property involves the Hessian of $H_0(J_1,J_2,\cdots, J_N)$, 
where $J_1, \cdots, J_N$ are actions.
The key issue is that the standard Kolmogorov-Arnold-Moser (KAM) theorem\cite{Lichtenberg}, 
as well as many other powerful mathematical results that determine the fate of integrable Hamiltonians, 
assume that the Hamiltonian is not degenerate. 
However, as originally discussed in Ref. \onlinecite{delcastillo96} there are important, 
physically relevant problems described by degenerate Hamiltonian systems 
including transport in non-monotonic shear flows \cite{delcastillo2000}  
and magnetic chaos in reversed shear configurations \cite{Balescu98, Constantinescu2012}. 
In the latter case, the degeneracy is directly linked to the non-monotonicity of the $q$-profile. 
In particular, as it is well-known, in the Hamiltonian description of magnetic fields, 
the flux, $\chi$, plays the role of the action variable, 
and the unperturbed frequency, which corresponds to the poloidal transit frequency, 
is given by $\Omega_0=1/q(\chi)$. 
In this case, the destruction of shearless KAM surfaces, 
which corresponding to the destruction of ITB located where $q$ exhibits a minimum, 
is a problem outside the range of applicability of the standard KAM theory, 
and extended theories for non-twist (degenerate) systems\cite{delcastillo97, Morrison2009, delsham2000, gonzalez2014} are necessary.
The fate of shearless KAM tori in generic degenerate Hamiltonian systems was first studied in Ref. \onlinecite{delcastillo96},
where its was shown that there are two processes at play. 
One is separatrix reconnection that involves a change in the topology of the iso-contours of the Hamiltonian 
and the other is the remarkable resilience of shearless tori due to 
the anomalous scaling of the higher-order islands forming in the vicinity of the shearless point 
(e.g., minimum of the $q$-profile). 
These two processes, separatrix reconnection and resilience of shearless KAM curves, 
have been studied in significant detail in the context of the dynamics of magnetic field line orbits. 
However, to the best of our knowledge, 
the present paper is the first systematic study of the impact of these processes in the full particle orbits.  
In particular, we are interested in the study of how the topology of the full orbits changes 
as a result of the changes in the magnetic field topology in reversed shear configurations. 
Also, we are interested in the study the relationship 
between the robustness of magnetic flux surfaces near the minimum of the $q$-profile (i.e. magnetic field ITBs) 
and the robustness of full particle orbits transport barriers (i.e. particles ITBs).

Given these aforementioned facts, the paper consists in two different specific parts.
In the first part, we investigate the motion of particles in a configuration
when field lines are integrable. In this setting the Poincar\'e plots
are taken on the iso-(angular) momentum set. 
One consequence of the existence of chaotic region is that 
there is no global third integral corresponding to the magnetic moment.
It is thus important to look into this kind of chaos for checking validity of 
global gyrokinetic reductions and  
for estimation of error coming from the reductions.

In the second part, 
we focus on the relation between the magnetic field line and the particle trajectory, 
which are closely related. 
Roughly speaking, the ``zeroth'' approximation of the guiding center reduction is along the magnetic
field line. It is thus worthwhile to compare both the particle trajectory and the magnetic field line, 
when the particle trajectory is chaotic while the magnetic field line is regular. 
When the magnetic field lines become chaotic this issue remains, 
however in this setting the Poincar\'e sections on the iso-$\mu$ set or iso-angular momentum sets are not suitable anymore, 
one reason being that for instance we cannot define 
a single common section to represent the particle trajectory and the magnetic field line. 
To circumvent this problem, we propose a method using some periodic average of the trajectory
in the vicinity of the magnetic section. 
This method allows us to compare the particle trajectory and the magnetic field line on the same Poincar\'e section. 
The result that we find is that in the regular field, 
we see a particle trajectory that is roughly regular, 
but the shape of the trajectory is completely different from the magnetic field line. 
Another peculiar effect that we pinpoint is that 
there exists a region in which ``regular'' particle trajectories are found 
but the magnetic field line is chaotic.
This means that an ITB can be created in a chaotic magnetic field 
at least for a certain population of particles with some given energies. 
To be more specific, we find that the particles can move in the chaotic magnetic region almost randomly, 
but they cannot cross some specific region of phase space, 
they are trapped in a given region implying that an ITB exists.

In order to introduce jointly the two considered questions we have organized the paper as follows. 
First, in Sec. \ref{sec:model}, we introduce the setting that  we deal with. 
The reduced single particle  Hamiltonian for the unperturbed integrable part is derived, 
and we used a more general formalism than what was proposed in Ref. \onlinecite{Cambon2014}.
Then, in Sec. \ref{sec:poincare}, we introduce and detail the two different methods in order to visualize the particles motion 
and compute numerically  Poincar\'e sections, for the two considered problems.  
We then move on to the actual obtained results in Sec. \ref{sec:motion} and discuss the motion of  the charged particle. 
More specifically, in Sec. \ref{sec:chaos-in-regular}, the existence of chaotic motion in a regular magnetic field  is presented, 
and in Sec. \ref{sec:regular-in-chaos}, 
the motion of the particles in a chaotic magnetic field and the problems related to the creation of ITB are considered. 
Finally we conclude.

\section{Model\label{sec:model}}

\begin{figure}[tb]
\begin{centering}
\includegraphics[width=7cm]{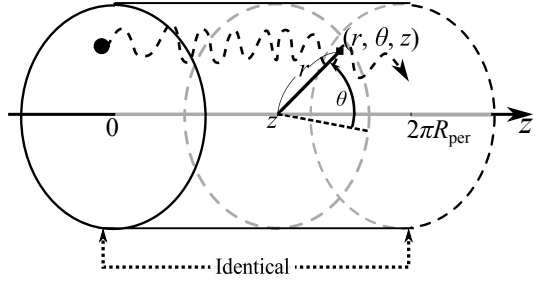}
\par\end{centering}
\caption{Sketch of the considered magnetic configuration and coordinate system.\label{fig:Sketch-of-the}}
\end{figure}

In this section we introduce the relevant parameters of the considered system and how we have dealt with it. 
As mentioned before, we consider the motion of a charged particle in a given static magnetic field $\mathbf{B}$.
The geometry is a cylinder of radius $r_{\rm cyl}$ and a periodicity of $2\pi R_{{\rm per}}$ along the axis 
(see Fig.~\ref{fig:Sketch-of-the}),
where $R_{\rm per}$ is interpreted as ``toroidal radius'' of a flat torus.
The variables are normalized as done in Ref.~\onlinecite{Cambon2014} by 
$\tilde{\mathbf{r}} =\mathbf{r}/r_{{\rm cyl}}$,
$\tilde{t} =\omega_{{\rm gyr}}t$, $\omega_{{\rm gyr}}=|e|B_{0}/M$, 
and 
$\tilde{\mathbf{v}} ={\rm d}\tilde{\mathbf{r}}/{\rm d}\tilde{t}$, 
where $r_{{\rm cyl}}$ is radius of the cylindrical domain, and $e$,
$M$ are the charge and mass of the particle. 
We use the normalized kinetic energy $\tilde{H}=\tilde{\mathbf{v}}^{2}/2$. 
To obtain the value of the energy in  keV, we simply transform 
\begin{equation}
	H=\frac{M\|\mathbf{v}\|^{2}}{2}=M\omega_{{\rm gyr}}^{2}r_{{\rm cyl}}^{2}\tilde{H}=\frac{e^{2}B_{0}^{2}r_{{\rm cyl}}^{2}}{M}\tilde{H}
\end{equation}
Then the keV values of energy  are obtained by just multiplying $(9.58\times10^{4})B_{0}^{2}r_{{\rm cyl}}^{2} \sim 10^5$,
$(9.64\times10^{4})B_{0}^{2}r_{{\rm cyl}}^{2}  \sim 10^5$, or $(1.76.\times10^{8})B_{0}^{2}r_{{\rm cyl}}^{2}  \sim 10^8$,
to the dimensionless energy $\tilde{H}$, for the proton, the alpha particle, or the electron respectively. 
In what follows, we omit tildes for normalized variables.

The magnetic field $\mathbf{B}$ is defined using the vector potential
$\mathbf{A}$ by $\mathbf{B}=\nabla\wedge\mathbf{A}$. Let $(r,\theta,z)$
be the cylindrical coordinates, which are defined as $(x,y)=(r\cos\theta,r\sin\theta)$,
and the $z$-axis coincides with the cylindrical axis. 
Let $q(r)$ be a safety factor (winding number) and define the function $F(r)$ 
\begin{equation}
F(r)=\frac{B_{0}}{R_{{\rm per}}}\int^{r}\frac{r'}{q(r')}{\rm d}r'\:.\label{eq:F_r}
\end{equation}
We divide the vector potential $\mathbf{A}$ into an unperturbed integrable part and a perturbation, 
\begin{equation*}
\mathbf{A}=\mathbf{A}_{0}+\epsilon\mathbf{A}_{1}\:,
\end{equation*}
where 
\begin{equation}
	\label{eq:A0}
	\begin{split}
		\mathbf{A}_{0}(r) 
		&=-\frac{B_{0}y}{2}\mathbf{e}_{x}+\frac{B_{0}x}{2}\mathbf{e}_{y}-F(r)\mathbf{e}_{z}\\
		&=-\frac{B_0 r}{2} \mathbf{e}_\theta-F(r)\mathbf{e}_{z}
	\end{split}
	\:,
\end{equation}
and 
\begin{equation}
	\mathbf{A}_{1}(r,\theta,z)
	=\sum_{m,n}\hat{A}_{1}^{mn}(r)\cos\left(m\theta-nz/R_{{\rm per}}\right)\mathbf{e}_{z}\:.
\end{equation}
Here $\epsilon$ is a small parameter, and \textbf{$\mathbf{e}_{i}$} is the base unit vector for $i=x,y,z,r,\theta,...\textrm{etc.}$.

In order to obtain the magnetic field lines, 
we recall briefly that these lines are governed by a Hamiltonian system\cite{Cary83, Abdullaev2013}. 
Let $\chi$ denote the poloidal magnetic flux of the unperturbed part, we have $\chi\equiv B_{0}r^{2}/2$. 
We can associate $\chi$ and the poloidal angle $\theta$ as canonically conjugated variables, and
the $z$ coordinates acts as the ``time.'' 
We then get the equation of the field line given by 
\begin{equation}
	\label{eq:field_line}
	\begin{split}
		\frac{{\rm d}\chi}{{\rm d}z} 
		& =-\frac{\partial\mathcal{H}}{\partial\theta},\quad 
		\frac{{\rm d}\theta}{{\rm d}z} 
		 =\frac{\partial\mathcal{H}}{\partial\chi},
	\end{split}
\end{equation}
where $\mathcal{H}=-A_{z}(\theta,\chi,z)$ . 
We recall that in the zeroth order approximation, a charged particle moves along the field line on average.

\subsection{The model of a particle moving in magnetic field configuration\label{sec:particle-model}}

The dynamics of a charged particle moving in a magnetic field is governed by the Hamiltonian  
\begin{equation}
H(\mathbf{q},\mathbf{p})=\frac{\|\mathbf{p}-\mathbf{A}(\mathbf{q})\|^{2}}{2}\label{eq:Ham}
\end{equation}
where we have normalized the mass and the charge of the particle to 1,
and $\|\bullet\|$ denotes the Euclidean norm in $\mathbb{R}^{3}$ space. 
Taking the Legendre transformation, we have the Lagrangian,
\begin{equation}
L=\frac{\dot{\mathbf{q}}^{2}}{2}+\dot{\mathbf{q}}\cdot\mathbf{A}(\mathbf{q})
\end{equation}
where the upper-dots denote the time derivative, ${\rm d}/{\rm d}t$, 
and the center-dot denotes the scalar-product in $\mathbb{R}^{3}$. 
For the considered magnetic configuration as 
\begin{equation}
	\label{eq:L}
	\begin{split}
 		L
 		& =\frac{\left(\dot{r}^{2}+r^{2}\dot{\theta}^{2}+\dot{z}^{2}\right)}{2}+\frac{B_{0}}{2}r^{2}\dot{\theta}+\dot{z}A_{z}(r,\theta,z)\\
	 	& =\frac{\left(\dot{r}^{2}+r^{2}\dot{\theta}^{2}+\dot{z}^{2}\right)}{2}+\frac{B_{0}}{2}r^{2}\dot{\theta}-\dot{z}F(r)\\
	 	& \quad+\epsilon\dot{z}\sum_{m,n}\hat{A}_{1}^{mn}(r)\cos\left(m\theta-nz/R_{{\rm per}}\right).
	\end{split}
\end{equation}
In what follows we examine the invariants of the motion for two cases; 
unperturbed motion $(\epsilon=0)$, and perturbation with one mode, i.e. 
$\hat{A}_{1}^{mn}(r)\neq0$ for only one $(m,n)$. 
Perturbations involving several modes will be addressed in a future paper.

\subsection{Unperturbed motion\label{sec:unperturbed}}

When the parameter $\epsilon=0$, $\theta$ and $z$ are cyclic and the conjugate momenta, 
the angular momentum around the $z$-axis,
\begin{equation}
	p_{\theta}=\frac{\partial L}{\partial\dot{\theta}}=r^{2}\dot{\theta}+\frac{B_{0}r^{2}}{2},
	\label{eq:p_theta}
\end{equation}
and the translation momentum along the $z$-axis, 
\begin{equation}
	p_{z}=\frac{\partial L}{\partial\dot{z}}=\dot{z}+A_{0}^{z}=\dot{z}-F(r)\label{eq:p_z}
\end{equation}
are invariants. 
Further, due to the time translational symmetry, the energy function $H$ is also an invariant of motion. 
Thus, there exist three integrals of motion and the unperturbed motion is completely integrable. 
Let us note as well that the field lines governed by Eq.~\eqref{eq:field_line} are also integrable.

Taking into account the invariants $p_{\theta}$ and $p_{z}$,
we can build an effective Hamiltonian with one degree of freedom. 
By making use of the cylindrical coordinates, the kinetic energy $H$ is written as 
\begin{equation}
	H=\frac{\left(\dot{r}^{2}+r^{2}\dot{\theta}^{2}+\dot{z}^{2}\right)}{2}.
\end{equation}
Substituting Eqs.~\eqref{eq:p_theta} and \eqref{eq:p_z} and $p_{r}=\partial L/\partial\dot{r}=\dot{r}$,
we end up with the desired one-degree of freedom effective Hamiltonian $H_{{\rm eff}}$
\begin{equation}
	\begin{split}
		H = H_{{\rm eff}}(r,p_{r}) 
		& =\frac{p_{r}^{2}}{2}+V_{{\rm eff}}(r;p_{\theta},p_{z}),
		\\
		V_{{\rm eff}}(r;p_{\theta},p_{z}) 
		& =\frac{p_{\theta}^{2}}{2r^{2}}+\frac{B_{0}r^{2}}{8}+\frac{(F(r)+p_{z})^{2}}{2}  -\frac{B_{0}p_{\theta}}{2}, 
	\end{split}
	\label{eq:H_eff}
\end{equation}
where constants $(p_{\theta},p_{z})$ are given by the initial conditions.

\subsection{Motion with single mode perturbation\label{sec:single-mode}}

We now consider the motion of a charged particle in the perturbed field when only one mode is presented. 
This means, we are considering the case $\epsilon>0$ and $\hat{A}_{1}^{mn}(r)\neq0$ for a pair of $(m,n)$, 
and $\hat{A}_{1}^{mn}(r)=0$ for other pairs of $(m,n)$. 
In this situation the two translational and rotational symmetries are broken 
and the two associated momentums $p_{\theta}$ and $p_{z}$ are no longer constants of the motion. 
However, this is an invariant associated with the helical symmetry along the curve 
$m\theta-nz/R_{{\rm per}} = \mathrm{Const.}$.
To exhibit it, we introduce new variables $(R,\Theta,\zeta)$ given by
$R=r$, $\Theta=\theta-nz/mR_{{\rm per}}$, and $\zeta=z$, 
we rewrite the Lagrangian \eqref{eq:L} as 
\begin{equation}
	\begin{split}
		L= & \frac{1}{2}\left[\dot{R}^{2}+R^{2}\left(\dot{\Theta}+\frac{n\dot{\zeta}}{mR_{{\rm per}}}\right)^{2}+\dot{\zeta}^{2}\right]\\
		&+\frac{B_{0}}{2}R^{2}\dot{\Theta}
		+\dot{\zeta}\left(\frac{nB_{0}R^{2}}{2mR_{{\rm per}}}+A_{\zeta}(R,\Theta)\right).
	\end{split}
\end{equation}
The coordinate $\zeta$ is now cyclic, and 
\begin{equation}
	p_{\zeta}:=\frac{\partial L}{\partial\dot{\zeta}}=\frac{n}{mR_{{\rm per}}}p_{\theta}+p_{z}
\end{equation}
is an integral of motion \cite{Landau-mechanics}. 
In this case, there are only two integrals of motion (the kinetic energy is of course still an invariant),
and as a consequence it is possible that the system is non-integrable.
If it is the case, we can expect the presence of Hamiltonian chaos, as will be shown later.

Regarding the field lines, they remain integrable. 
Indeed looking at Eq.~\eqref{eq:field_line}, the change of variable 
$\theta\mapsto\Theta+nz/mR_{\rm per}$ 
corresponds to a Galilean transformation which transforms the non autonomous equation into an autonomous one. 
Then, changing to the coordinates $(\Theta,\chi)$, 
we end up with an autonomous Hamiltonian system  with one degree of freedom, which is completely integrable.

\subsection{Reversed shear magnetic field \label{sec:detail}}

Non-monotonic $q$-profiles are known to create magnetic ITB \cite{Constantinescu2012, Balescu98}
Since we are interested in relationship between magnetic ITB and an effective particle barrier, 
we consider a magnetic configuration and  safety factor $q(r)$, that was already proposed in \cite{Blazevski2013}: 
\begin{equation}
	\label{eq:safety}
	q(r)=q_{0}\left[1+\lambda^{2}\left(r-\alpha\right)^{2}\right],
\end{equation}
where $q_{0},\alpha$, and $\lambda$ are some constants.
Note that because $q$ is not monotonic, there can be two resonant magnetic surfaces for a given rational $q$. 
The function $F(r)$ corresponding to this $q(r)$ is  
\begin{equation}
	\begin{split}
		F(r)= & \frac{B_{0}}{R_{{\rm per}}}\int_{0}^{r}\frac{r'}{q(r')}{\rm d}r'\\
		= & \frac{1}{R_{{\rm per}}q_{0}}\left[\frac{\ln\left(1+\lambda^{2}\left(r-\alpha\right)^{2}\right)}{2\lambda^{2}}+
		\frac{\arctan\left(\lambda(r-\alpha)\right)}{\lambda}\right].
	\end{split}
\end{equation}
The amplitudes of Fourier modes of the perturbation term are set up as follows. 
Following Ref.~\onlinecite{Blazevski2013}, 
we choose the amplitudes $\hat{A}_{1}^{mn}(r)$ of the Fourier modes 
so that $A_1$ has peaks (maximum intensities) at the points $r_{\pm}^{\ast}$ satisfying
$q(r_{\pm}^{\ast})=m/n$ (i.e on a resonant surface), 
localized at
$r_{\pm}^{\ast}=\alpha\pm\frac{1}{\lambda}\sqrt{\frac{m}{nq_{0}}-1}$
and write 
\begin{equation}
\hat{A}_{1}^{mn}(r)=a(r)\left(A_{+}^{mn}(r)+A_{-}^{mn}(r)\right),
\end{equation}
where 
\begin{equation}
	\begin{split}
		\hat{A}_{\pm}^{mn}(r) & =
		\left(\frac{r}{r_{\pm}^{\ast}}\right)^{m}\exp\left(-\frac{(r-r_{\pm}^{0})^{2}+(r_{\pm}^{0}-r_{\pm}^{\ast})^{2}}{2\sigma^{2}}\right)\\
		r_{\pm}^{0} & =r_{\pm}^{\ast}-m\sigma^{2}/r_{\pm}^{\ast},
	\end{split}
\end{equation}
where $l$ and $\sigma$ are positive constants. 
The function $a(r)$ is written as 
\begin{equation}
	a(r)=\frac{1}{2}\left(1-\tanh\left(\frac{r-1}{l}\right)\right),
\end{equation}
which allows to kill the perturbation outside of the cylinder ($r>1$).

\section{Poincar\'e section\label{sec:poincare}}

Since the trajectory lies in a six dimensional phase space, 
it is difficult to clearly visualize the full trajectory. 
In order to circumvent this problem, 
we make Poincar\'e plots on a two dimensional subset,
which is relatively simple for the field lines, but is however rather complex for the full particle orbit. 
Indeed, one could naively consider identical section for the field lines and the particles, 
meaning section on the planes $z=2\pi R_{{\rm per}}N$ for $N\in\mathbb{N}$. 
However, due to the Larmor gyration, the particle trajectory becomes thick even when the particle motion is completely integrable. 
This Larmor gyration blurs the details of the particle motion and makes these more or less useless. 
In what follows, we discuss two methods to ``suppress'' the thickness of the trajectory and alleviate the problem.

\subsection{On the invariant set for unperturbed integral system\label{sec:iso}}

As mentioned before, without perturbation, the motion is integrable. 
However, when only one mode is present, only two constants of motion exist. 
In this setting, we choose the Poincar\'e section on the invariant sets of the variables 
which are integrals in the unperturbed system but not in the perturbed system, for example,
the angular momentum around $z$-axis $p_{\theta}$, the momentum $p_{z}$, 
the effective Hamiltonian $H_{{\rm eff}}$ [Eq. \eqref{eq:H_eff}],
and the unperturbed part of the Hamiltonian, $H_{0}=H_{{\rm eff}}$.
In this paper we settled on using the iso-$p_{\theta}$ plane. 
As will be shown, this method is successful in visualizing the presence of chaos generated 
by the slow motion of the separatrix in the $(r,p_{r})$ phase space. 
However, it is then difficult to relate it to the field lines 
and we cannot compare the charged particle motion with the magnetic field line profile using this strategy.

\subsection{One period averaging method\label{sec:ave}}

To visualize the dynamics, even in the absence of constants of motions (except for the energy), 
we resort to use an averaging like method to perform a section. 
As will be shown this allows us to get rid of the ``blurriness'' induced by the Larmor gyration.  
However, we anticipate that the method may have some caveats if one considers extremely long time, 
and when slow adiabatic chaos is present.
Despite this, this method allows us to get a good idea of the regularity of the motion, 
at least for the considered simulation time, 
and none of the techniques used in the context of the regular magnetic field proved to be adequate.
One other advantage is that this allows a more visual comparison between particle motion and the magnetic configuration.

\begin{figure}[tb]
	\centering{}
	\includegraphics[width=7cm]{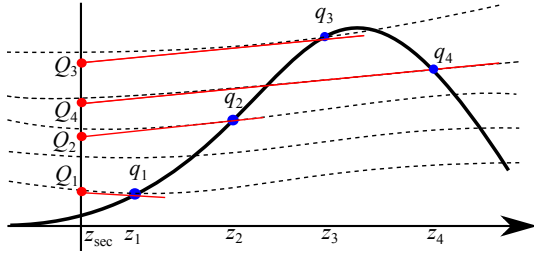} 
	\caption{ An illustrative explanation of $\mathbf{Q}_{k}$. The bold solid
	curve and broken curves are the particle's trajectory and the magnetic
	field curves, respectively. The red lines are tangent to the magnetic
	field curves for each $\mathbf{q}_{k}$. 
	This method assumes that the magnetic field is smooth 
	and that the particle does not move fast along the field lines.} 
	\label{fig:pp1} 
\end{figure}

As mentioned, in order to compare with magnetic field, 
we consider a similar section as the one performed for the field lines, i.e. 
we want to make a ``Poincar\'e'' section for each time when $z$ reaches $2\pi R_{{\rm per}}\mathbb{N}$. 
This will work for non energetic particles with small Larmor radii, 
however when the energy will be increased finite size Larmor effects will end up 
``blurring'' the section making it impossible to decipher wether or not the motion is regular or chaotic. 
This is why, we used the averaging technique.
In order to be more explicit, 
we present below the details of  the numerical procedure we used in order to perform this specific section.
Let us now detail the averaging procedure. We  note $\delta t$ the time step in the numerical integration.
The crossing of the section plane will occur  when  $z_{{\rm sec}}\in2\pi R_{{\rm per}}\mathbb{N}$, 
and thus around a time $t_n$ such that  $z(t_n)$ satisfies $z(t_n)<z_{{\rm sec}}<z(t_n+\delta t)$. 
It is within a time slice of this crossing that we perform a time average over one Larmor period.
The number of time steps $n_{{\rm L}}$ occurring during one Larmor gyration is approximately 
the integer part of $\tau_{\rm L}/ \delta t$, 
where $\tau_{{\rm L}}=2\pi/\|\mathbf{B}\|$ is the cyclotoronic period. 
We thus compute numerically  the  $n_{\rm L}$ points of the particle trajectory  $\mathbf{q}_{k}=\left(x(t_{k}),y(t_{k}),z(t_{k})\right)$ 
for $k=1,2,3,...,n_{{\rm L}}$, after crossing the section plane , meaning when $t_{k}=t_n+k\delta t$.
We then transport each point $q_k$ back on the section plane at $z_{{\rm sec}}$ along the magnetic field line that passes through $q_k$. 
We obtain then the corresponding images  $\{\mathbf{Q}_{k}\}_{k=1}^{n_{\rm L}}$ on the section $z = z_{\rm sec}$. 
In fact in order to compute numerically the points $\mathbf{Q}_{k}$, we do a first order approximation. 
We assume that the magnetic field $\mathbf{B}$ depends smoothly on $\mathbf{q}$, 
and that $z(t_{n_{\rm L}})$ is not so far from $z_{{\rm sec}}$. 
So to obtain the $\mathbf{Q}_{k}$, we simply  project $\mathbf{q}_{k}$ to the section along tangent lines of the magnetic field $\mathbf{B}$ at $\mathbf{q}_{k}$.
This is  shown schematically in Fig.~\ref{fig:pp1}. 
All the image points are on the section plane plane $z=z_{{\rm sec}}$, 
and we can write them in Cartesian coordinates as 
$\mathbf{Q}_{k}=\left(X_{k},Y_{k},z_{{\rm sec}}\right)$,
and the projection along the filed like yields numerically 
that $X_{k}$ and $Y_{k}$ are given respectively by 
\begin{equation}
	\label{eq:XY}
	\begin{split}
		X_{k} & =x(t_{k})-\frac{B_{x}(\mathbf{q}_{k})}{B_{z}(\mathbf{q}_{k})}(z(t_{k})-z_{{\rm sec}})\:,\\
		Y_{k} & =y(t_{k})-\frac{B_{y}(\mathbf{q}_{k})}{B_{z}(\mathbf{q}_{k})}(z(t_{k})-z_{{\rm sec}}).
	\end{split}
\end{equation}
To get the final position of our section we perform an average, 
meaning in the Poincar\'e sections of the particle motion that we present with the chaotic magnetic field,  
we plot the average points $\left(X(z_{{\rm sec}}),Y(z_{{\rm sec}})\right)$
that are obtained by  
\begin{equation}
	\left(X(z_{{\rm sec}}),Y(z_{{\rm sec}})\right)=
	\frac{1}{n_{{\rm L}}}\sum_{k=1}^{n_{{\rm L}}}(X_{k},Y_{k})\label{eq:XYave}
\end{equation}
on each plane $z=z_{{\rm sec}}$. 
As already mentioned using this average  we were able to transform a naively obtained ``blurry'' trajectory 
on the plane $z=z_{{\rm sec}}$ into a trajectory with a much better resolution 
(when motion is integrable or quasi-integrable on the considered time scales).
Roughly speaking we are making a pseudo-reduction: 
the equation of motion are rigorously numerically solved,  
but we perform an average similar to the gyro-average only when we take the Poincar\'e plot. 
The illustrative explanation is exhibited in Fig.~\ref{fig:pp1}.

\section{Charged particle motion in magnetic field\label{sec:motion}}

In this section we study 
the chaotic motion in an integrable magnetic field,  
and investigate the influence of the kinetic energy and initial pitch angle on 
the topology of particle trajectories.
In particular, we show that a ``regular'' region that acts
as an effective transport barrier can appear in a region in which
the magnetic field lines are chaotic.

Since we are computing long-time trajectories 
resolving the Larmor gyration, the equations of motion are integrated
using a sixth order implicit symplectic Gauss-Legendre method \cite{MacLchlan92}.
This is required in order to avoid non-Hamiltonian features such
as sinks and sources that may arise using, for instance, a standard Runge-Kutta method. 

\subsection{Hamiltonian particle chaos in integrable magnetic field\label{sec:chaos-in-regular}}

\begin{figure}[tb]
	\begin{centering}
		\includegraphics[width=7cm]{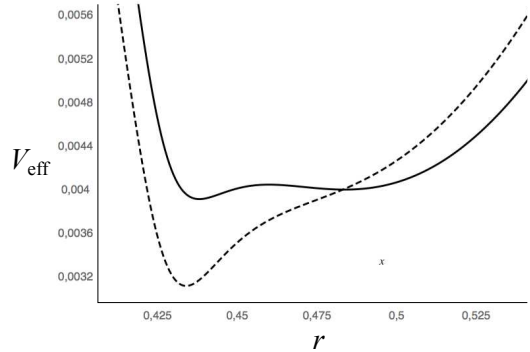}
	\end{centering}
	\caption{Illustration of an effective potential with one or two wells.
	The parameters are fixed as $q_0 = 0.12$, $\lambda = 55$, and $\alpha = \sqrt{0.18}$.
	Solid and broken curves are obtained for 
	$(p_\theta, p_z) = (0.13, 0)$ and $(0.125, 0.003)$ 
	respectively.
	}
	\label{fig:Effective_potential}
\end{figure}

\begin{figure}[tb]
	\begin{centering}
		\includegraphics[width=7cm]{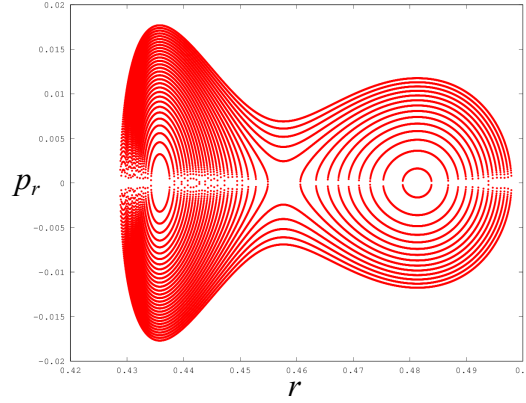}
	\end{centering}
	\caption{
		The contours of the effective Hamiltonian with a separatrix.
		Trajectories in phase space of the integrable case 
		are along each contour.
	}
\label{fig:Integrable_phase space}
\end{figure}

We first consider chaotic particle orbits in integrable magnetic field. 
As shown in Fig.~\ref{fig:Effective_potential}, depending on the values of the constants of the motions 
(initial conditions) $p_{\theta}$ and $p_{z}$ and 
a set of parameters $q_0$, $\lambda$ and $\alpha$ determining the safety factor \eqref{eq:safety}, 
the effective potential in Eq. \eqref{eq:H_eff} can have one or two minima. 
In the case of a double well as shown in Fig. ~\ref{fig:Integrable_phase space}, there is a separatrix.
The presence of this separatrix implies that a small perturbation
will lead to the formation of a stochastic layer and Hamiltonian chaos.
To investigate this possibility,  
we consider the magnetic field with the vector potential $\mathbf{A}=\mathbf{A}_{0}+\epsilon\mathbf{A}_{1}$
where $\mathbf{A}_{1}$ consists of a single cosine mode. 
As already mentioned, in this case, the magnetic field lines governed by Eq.~\eqref{eq:field_line} are integrable. 
We adjust the safety factor and the values of the invariants $p_{\theta}$ and $p_{z}$, 
so that $V_{{\rm eff}}$ leads to the presence of unstable (homoclinic)
point in the $(r,p_{r})$ phase space.
We take the Poincar\'e section for each iso-$p_{\theta}$ plane associated
with $\bar{p}_{\theta}:=T^{-1}\int_{0}^{T}p_{\theta}(t)dt$, $T=1000$.
We remind the reader that, once the perturbation is turned on, $p_{\theta}$
is not a constant of the motion anymore. 
We also mention that we can chose the Poincar\'e section as iso-$ap_{\theta}+bp_{z}$ plane
for any $a$ and $b$ holding $b/a\neq n/mR_{{\rm per}}$. 
These sections are equivalent when the perturbation consists of a single mode.
This is because, using the constant of the motion $p_{\zeta}$, it can
be rewritten as 
\begin{equation}
	\begin{split}
		ap_{\theta}+bp_{z} 
		& =ap_{\theta}+b\left(p_{\zeta}-\frac{np_{\theta}}{mR_{{\rm per}}}\right)\\
		 & =\left(a-\frac{nb}{mR_{{\rm per}}}\right)p_{\theta}+bp_{\zeta}.
	\end{split}
\end{equation}
\begin{figure}[tb]
\centering{}
	\begin{tabular}{cc}
	\begin{minipage}{0.45\hsize}
        	\begin{center}         
	{(a)}\\
	\includegraphics[width=7cm]{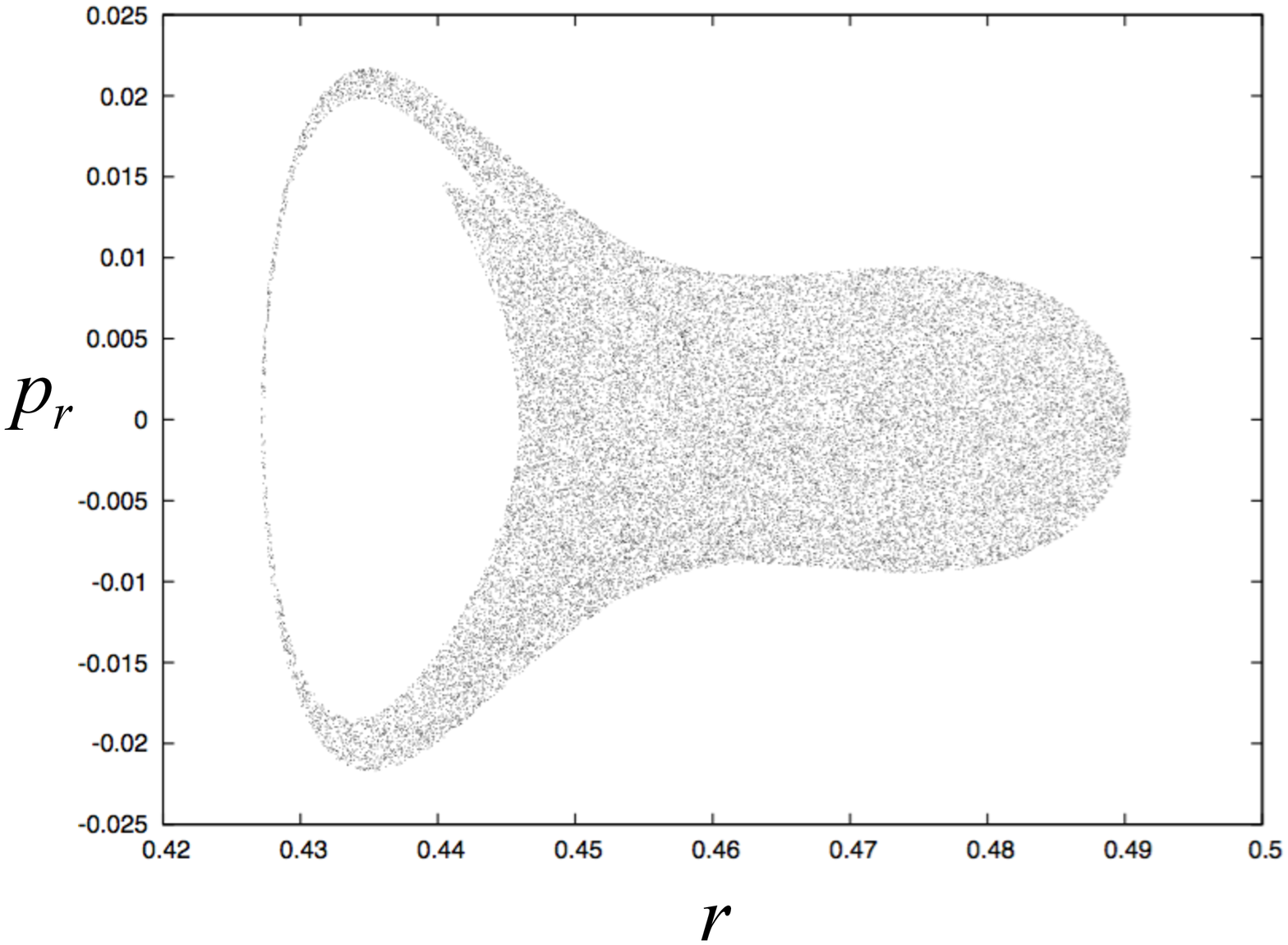}
	\end{center}
      	\end{minipage}
	\begin{minipage}{0.45\hsize}
        	\begin{center}         
	{(b)}\\
	\includegraphics[width=7cm]{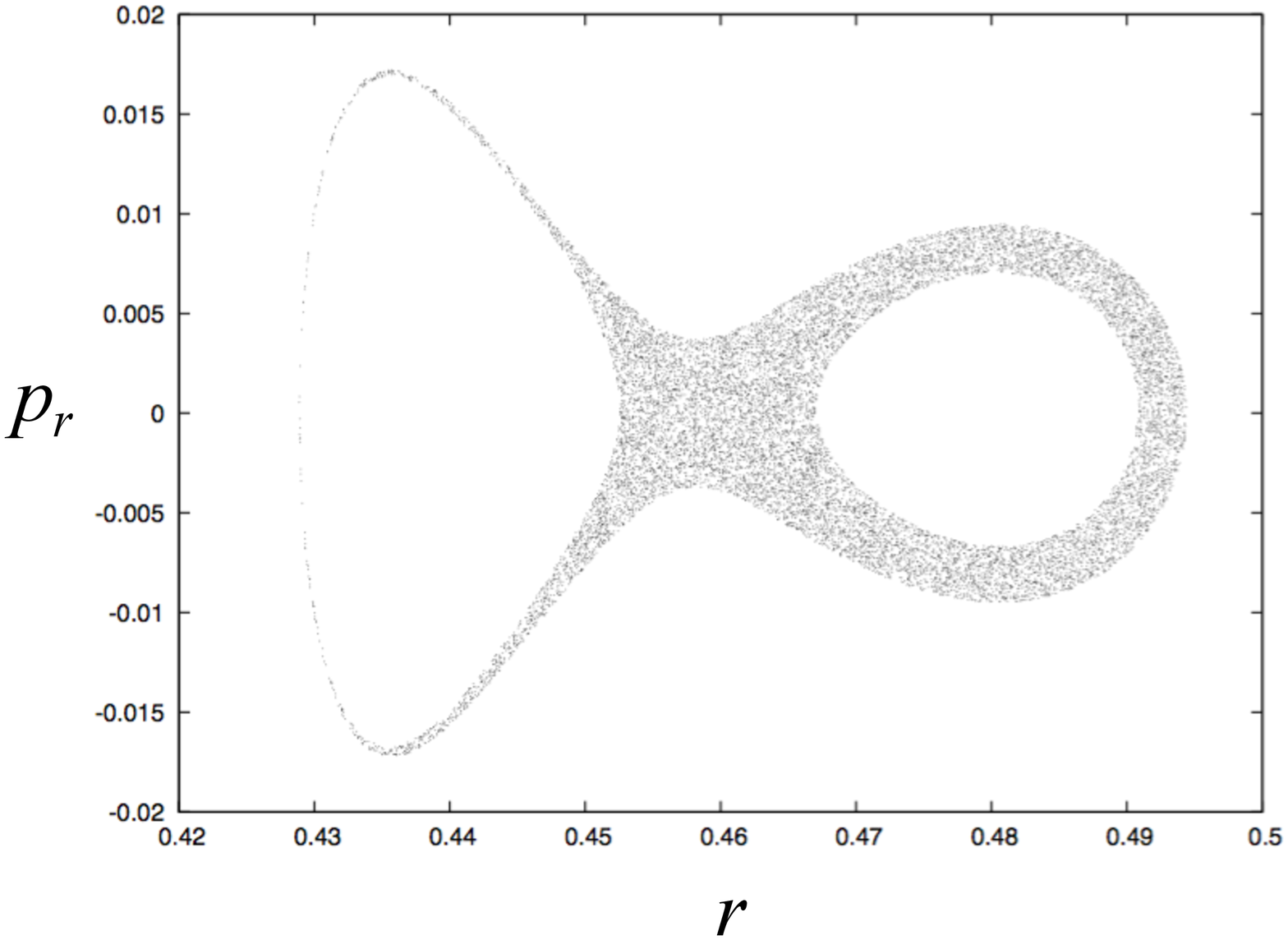}
		\end{center}
      	\end{minipage}
	\\
	\begin{minipage}{0.45\hsize}
        	\begin{center}         
	{(c)}\\
	\includegraphics[width=7cm]{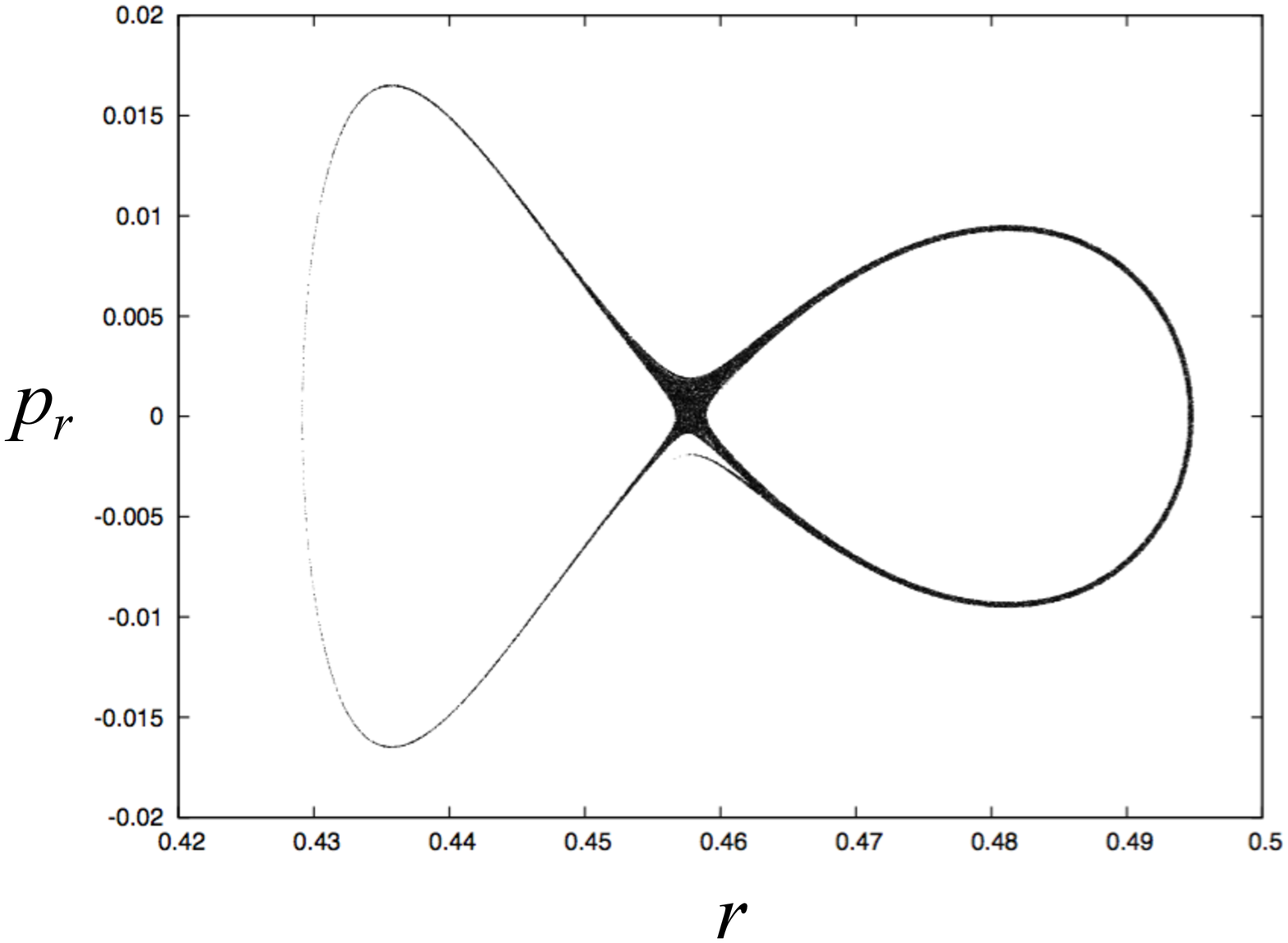}
	\end{center}
      	\end{minipage}
	\begin{minipage}{0.45\hsize}
        	\begin{center}         
	{(d)}\\
	\includegraphics[width=7cm]{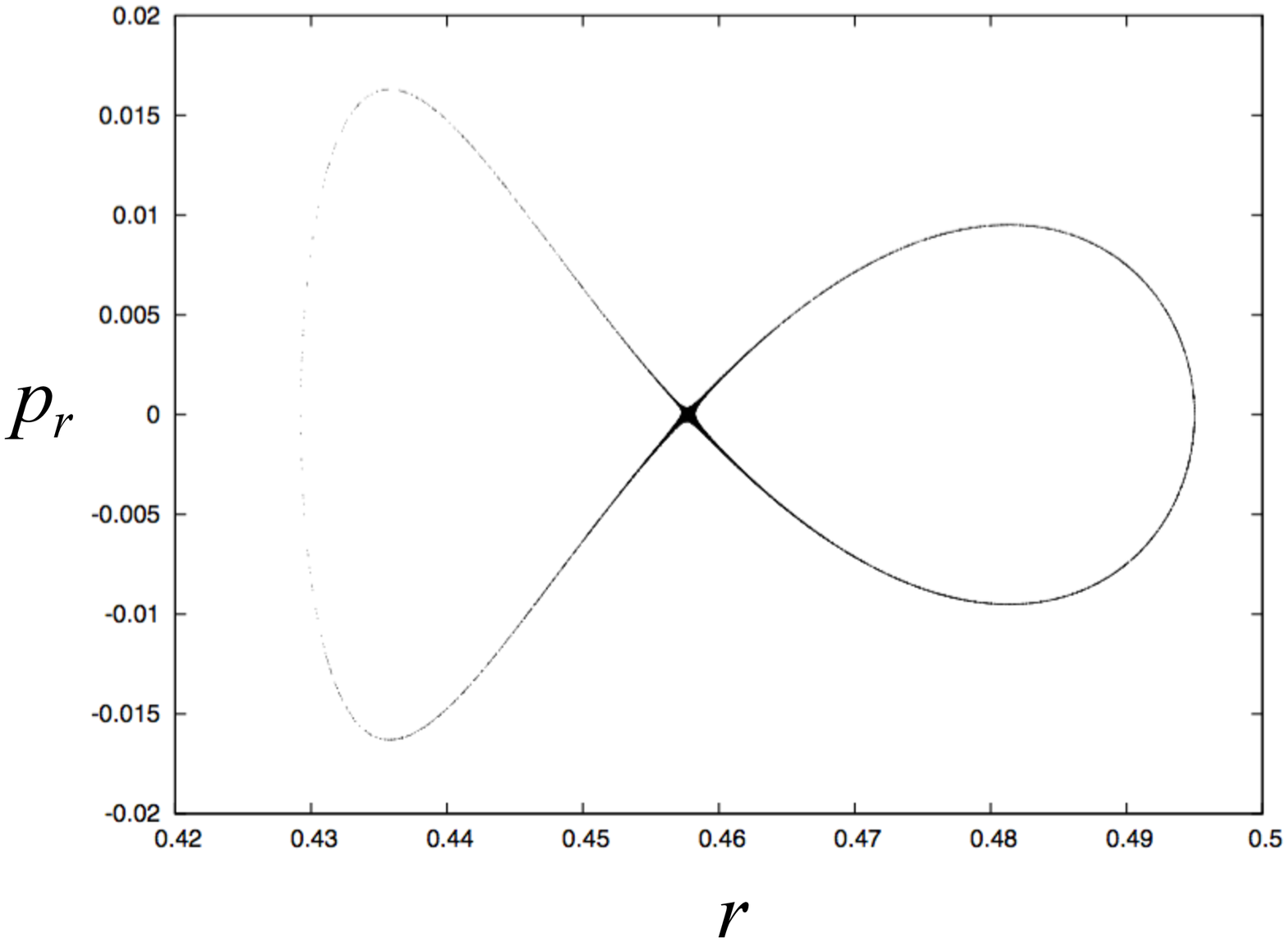}
	\end{center}
      	\end{minipage}
 \end{tabular}
	\caption{Poincar\'e plots on the iso-angular momentum planes for different perturbations, 
	$\epsilon=10^{-3},10^{-4},10^{-5}$, and $10^{-6}$. 
	The perturbation term consists of a single $(m,n)=(3,2)$ mode. 
	We set the unperturbed part as 
	$(p_{\theta},p_{z})=(0.13,0)$, $q_{0}=0.12$, $\lambda=55$, and $\alpha=\sqrt{0.18}$. 
	In this case, the effective Hamiltonian has a
	saddle point at $(r,p_{r})\simeq(0.4576,0)$. 
	The time step $\delta t=0.05$.}
	\label{fig:sepx_pp} 
\end{figure}

\begin{figure}[tb]
	\centering{}
	\includegraphics[width=8cm]{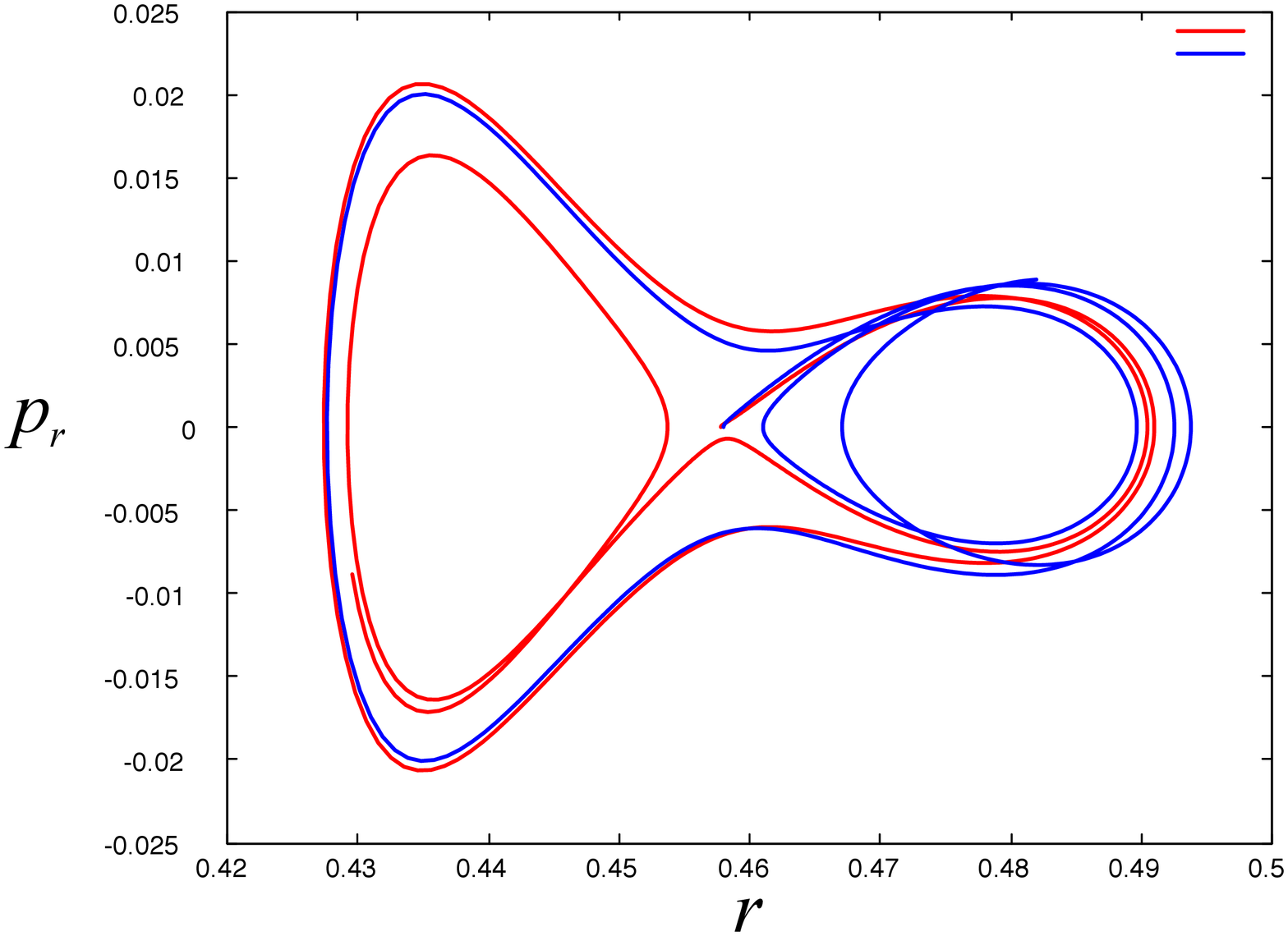} 
	\caption{
	Particle trajectories with
	the initial conditions $(x_{0},y_{0})=(0.4578,0),(0.4580,0)$. 
	The end points for each trajectory correspond to at 1000-th time step. We
	set the unperturbed part as $(p_{\theta},p_{z})=(0.13,0)$, $q_{0}=0.12$,
	$\lambda=55$, and $\alpha=\sqrt{0.18}$. In this case, the effective Hamiltonian
	has a saddle point at $(r,p_{r})\simeq(0.4576,0)$. The perturbation
	is set as $\epsilon=0.0001$, $(m,n)=(3,2)$. The time step $\delta t=0.05$. }
	\label{fig:sepx}
\end{figure}

The chaotic behavior exhibited in Fig.~\ref{fig:sepx_pp} 
results from the separatix crossing\cite{Neishtadt86,Tennyson86,Cary1986,Leoncini2009} shown in Fig.~\ref{fig:sepx}. 
Two trajectories are very close to each other at $t=0$, but they cross the separatrix at completely
different times, then the end point ($t=50$) are apart from each other.

The presence of chaos confirms the one exhibited by Cambon {\it et al.}\cite{Cambon2014}.
The existence of regions in phase space where the dynamics is not integrable 
when two constants of motion are preserved, hints strongly at the fact 
that no global third constant of the motion exists. 
It is thus likely that a constant of the motion related to the magnetic moment does not exist in these chaotic regions.
It seems therefore worthwhile to consider its effect when a global reduction such as gyrokinetics is performed, 
or at least to quantify the magnitude of the errors in the gyrokinetic theory induced 
by the existence of these regions in comparison to practical errors resulting from first order
development or the numerical scheme when we model the plasma using this theory. 
We do not deal with these issues in this paper. 
It should be remarked that, the particle showing chaos in the effective potential 
exhibited in Fig.~\ref{fig:Effective_potential} is energetic, 
and its kinetic energy is  approximately 400keV when this particle is an alpha particle or a proton.

We have shown so far that chaotic trajectories of charged particles
exist in a magnetic field with integrable field lines, we now 
investigate what happens when the field lines are chaotic.

\subsection{Regular motion in stochastic field\label{sec:regular-in-chaos}}

\begin{figure}[tb]
	\centering{}
	\begin{tabular}{cc}
	\begin{minipage}{0.45\hsize}
        			\begin{center}         
				{(a)}\\
	         			\includegraphics[width=7cm]{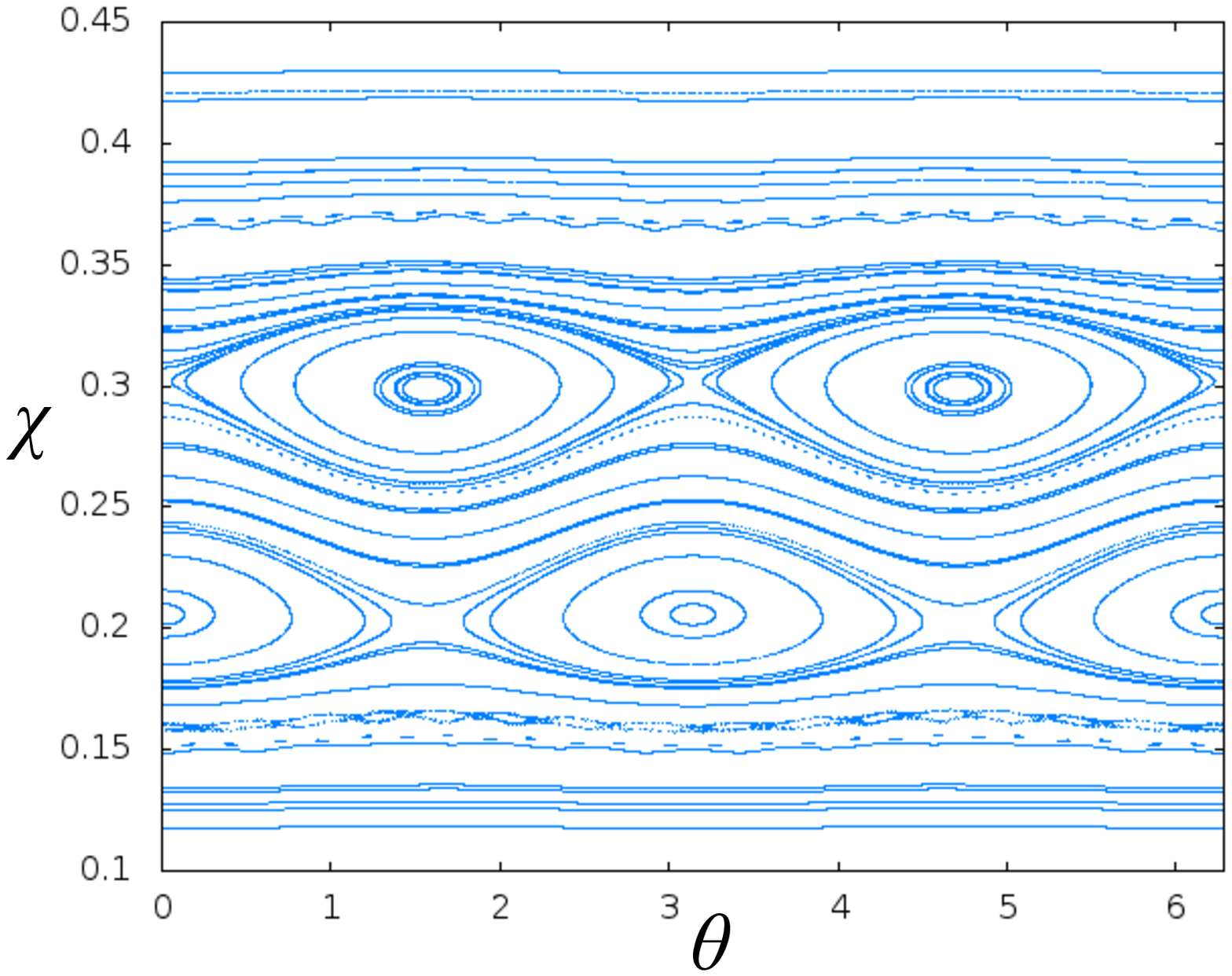}
        			\end{center}
      	\end{minipage}

	\begin{minipage}{0.45\hsize}
        			\begin{center}         
				{(b)}\\
	         			\includegraphics[width=7cm]{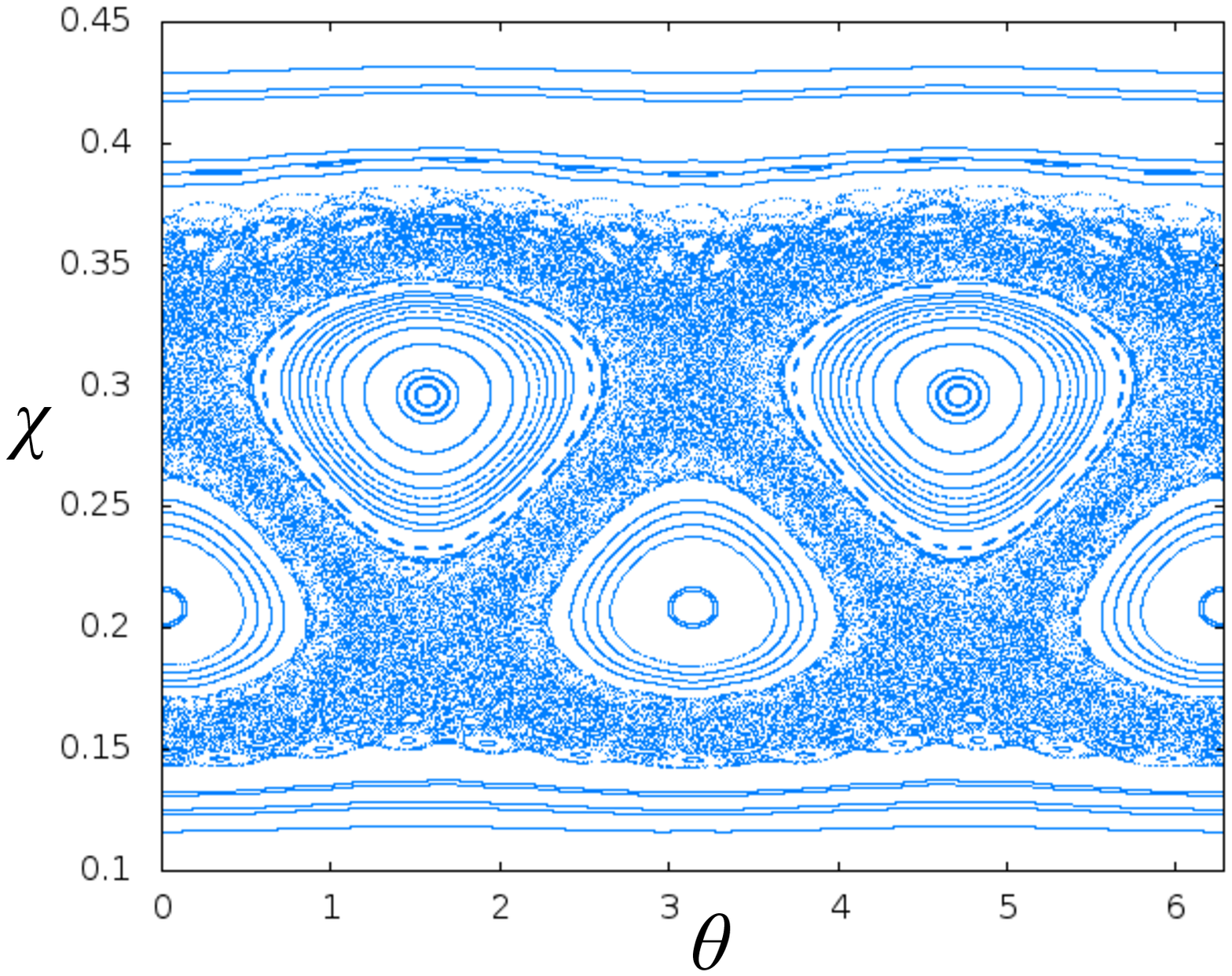}
        			\end{center}
      	\end{minipage}
\\
	\begin{minipage}{0.45\hsize}
        			\begin{center}         
				{(c)}\\
	         			\includegraphics[width=7cm]{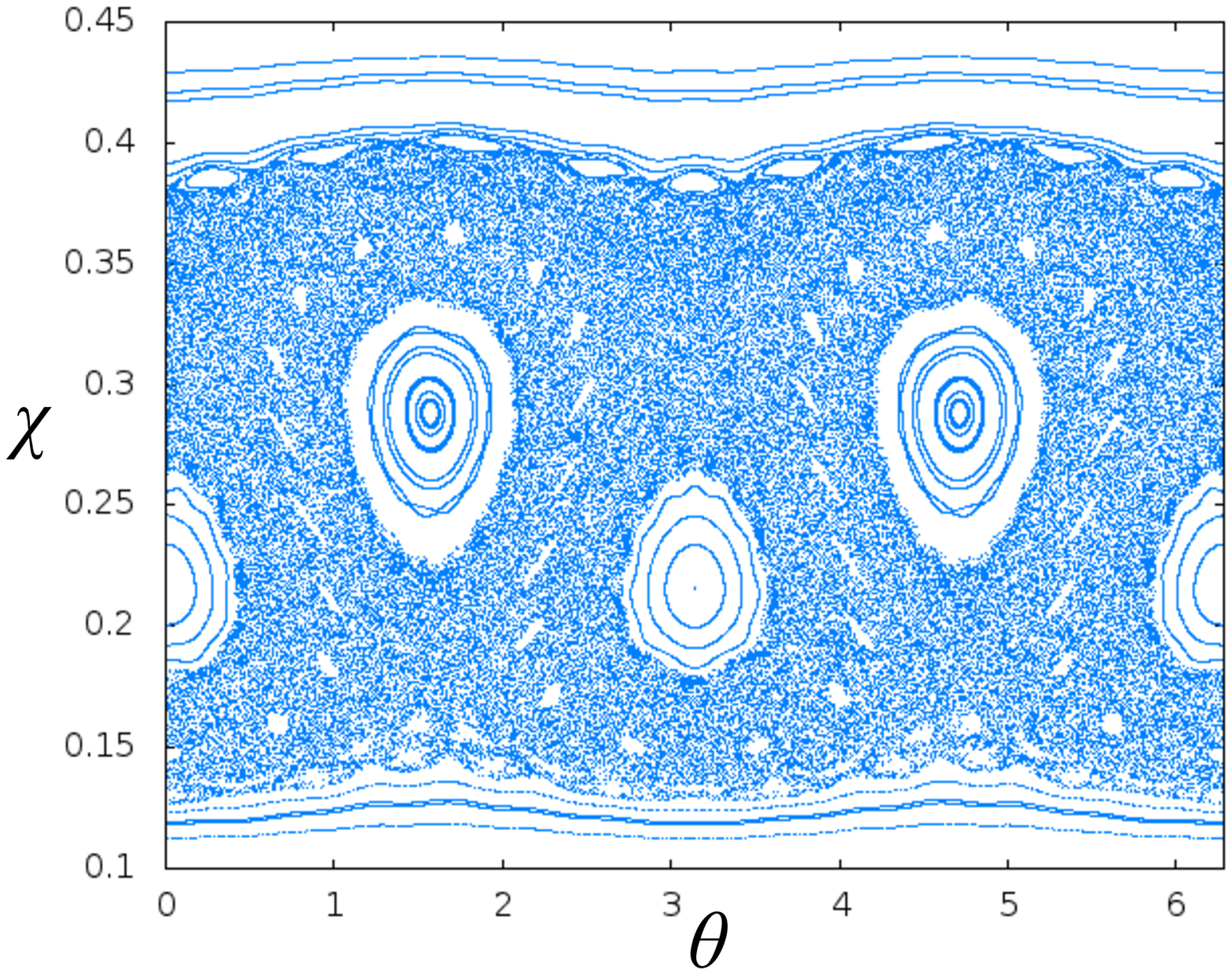}
        			\end{center}
      	\end{minipage}
  \end{tabular}

	\caption{Magnetic field lines with $\epsilon=0.0005,0.0015$, and $0.0045$, and $c=0.02$. }
	\label{fig:mag_c2e-2} 
\end{figure}

\begin{figure}[tb]
	\centering{}
	\begin{tabular}{cc}
	\begin{minipage}{0.45\hsize}
        			\begin{center}         
				{(a)}\\
	         			\includegraphics[width=7cm]{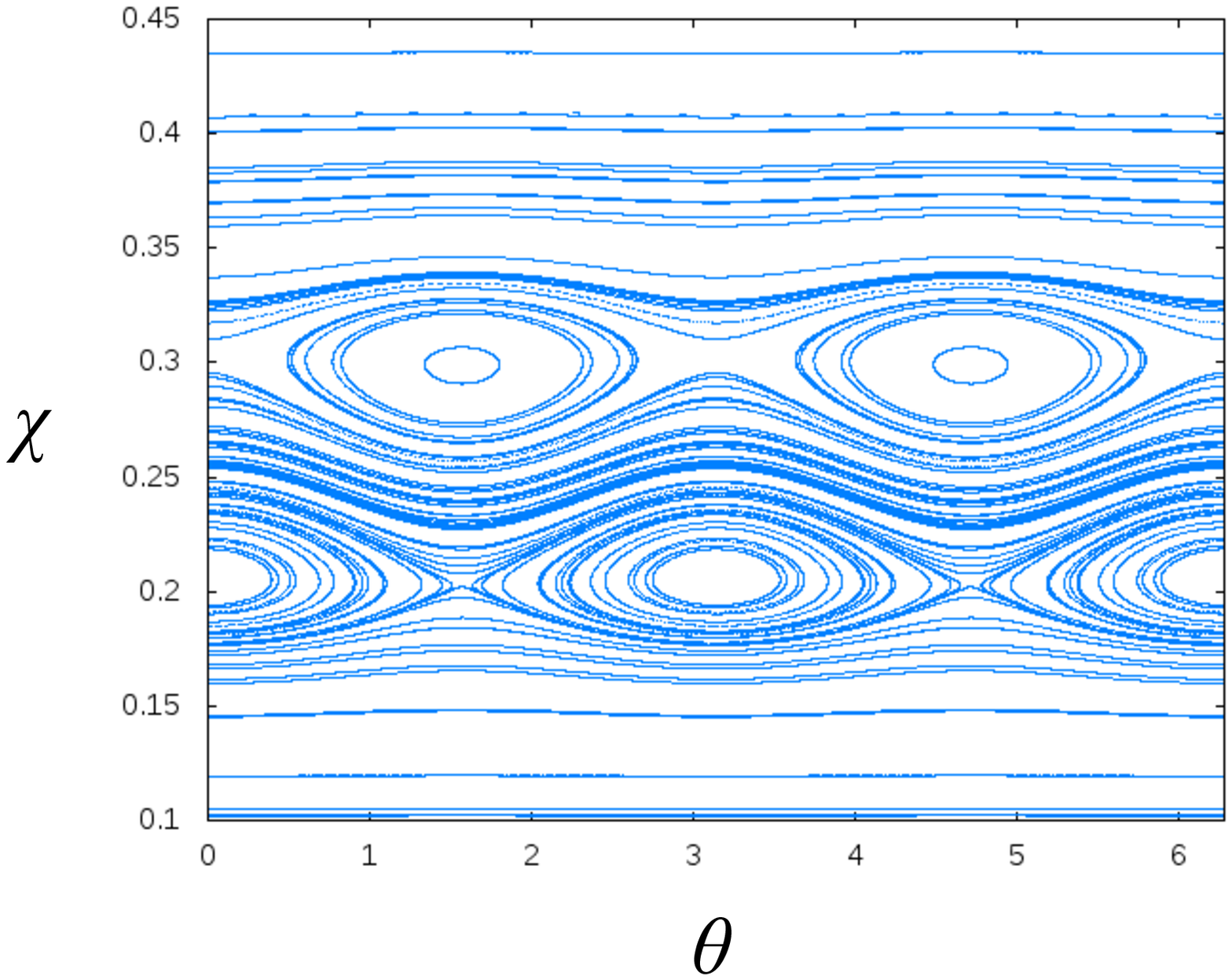}
        			\end{center}
      	\end{minipage}

	\begin{minipage}{0.45\hsize}
        			\begin{center}         
				{(b)}\\
	         			\includegraphics[width=7cm]{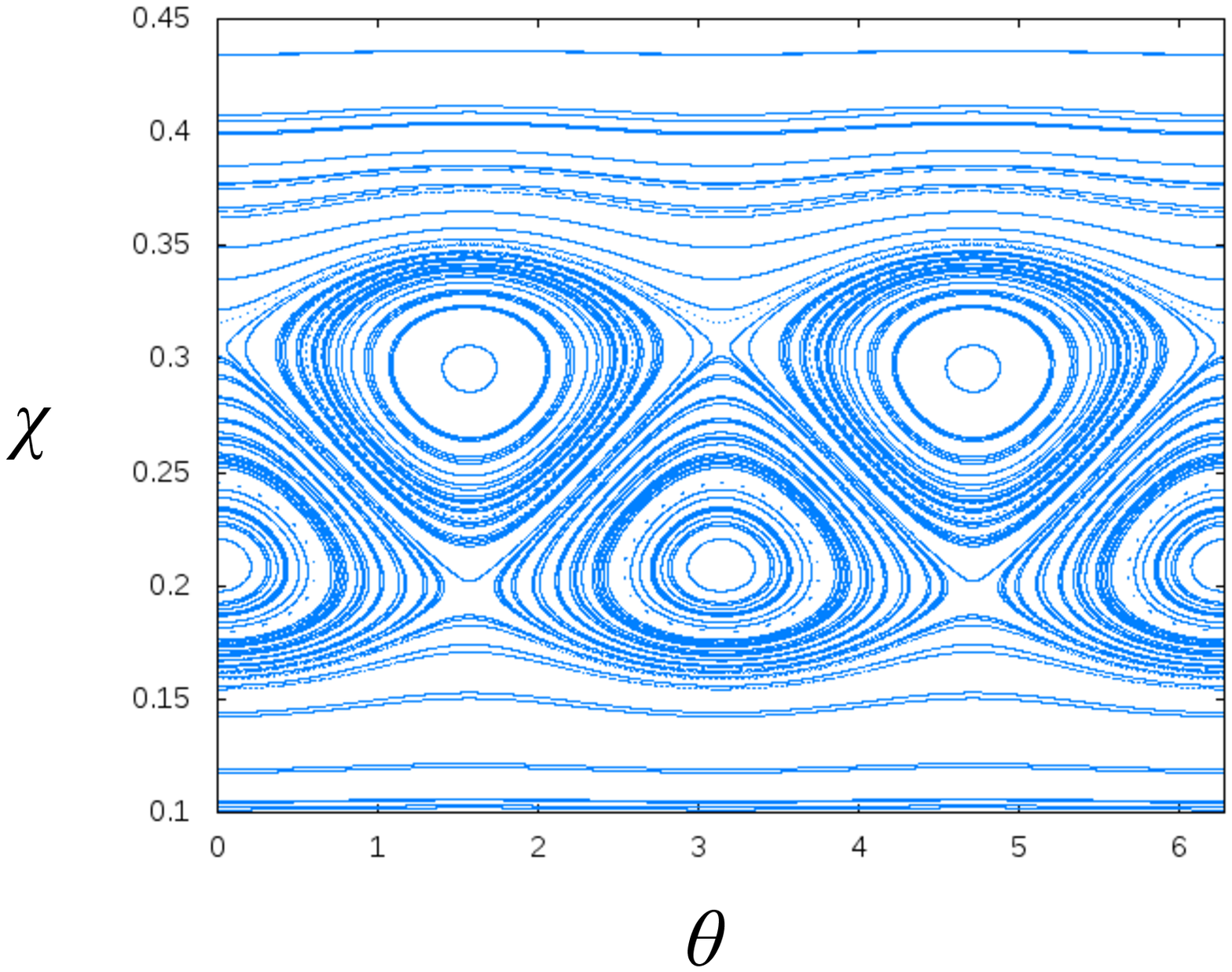}
        			\end{center}
      	\end{minipage}
\\
	\begin{minipage}{0.45\hsize}
        			\begin{center}         
				{(c)}\\
	         			\includegraphics[width=7cm]{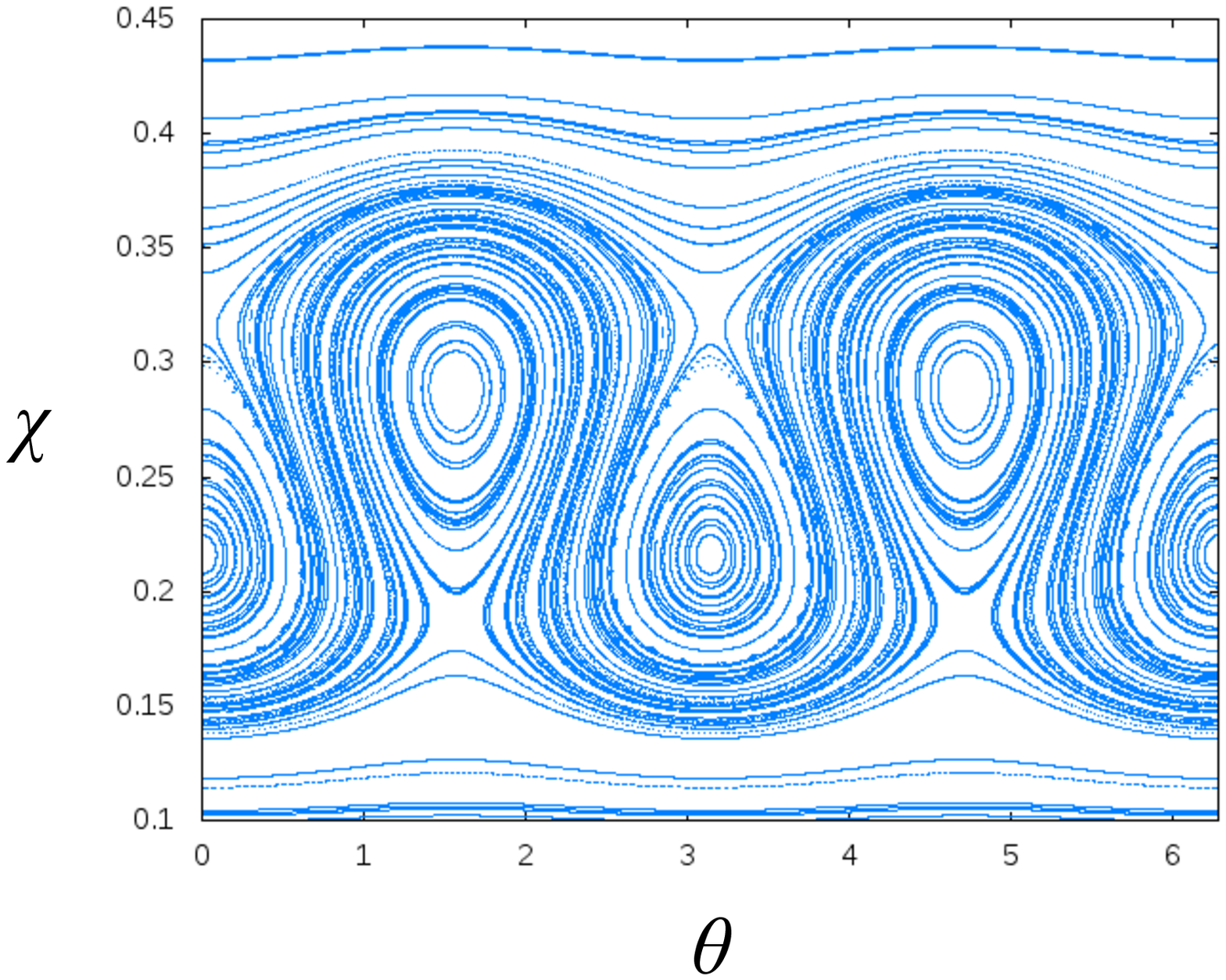}
        			\end{center}
      	\end{minipage}
  \end{tabular}

	\caption{Magnetic field lines, with $\epsilon=0.0005, 0.0015$, and $0.0045$ and $c=0$.}
	\label{fig:mag_c0} 
\end{figure}

In order to have a configuration with chaotic magnetic field lines,
we consider a magnetic perturbation with more than one mode. 
More specifically, we consider a perturbation of the type 
\begin{equation}
	\epsilon A_{1}=\epsilon\left(\hat{A}_{1}^{2\cdot3}(r)\cos(2\theta-3z)+c\hat{A}_{1}^{13\cdot17}(r)\cos(13\theta-17z)\right)
\end{equation}
The parameters in Eq. \eqref{eq:safety} are fixed:  $q_0 = 0.64$, $\lambda = 3$, and $\alpha = \sqrt{0.5}$.
Figures \ref{fig:mag_c2e-2} and \ref{fig:mag_c0} respectively show 
the Poincar\'e plots of the magnetic field lines for 
$\epsilon=0.0005,\:0.0015$, and $0.0005$ with $c=0.02$, 
and for $\epsilon=0.0005,\:0.0015$ and $0.0015$ with $c=0$. 
The other parameters are set as in the integrable case. 
The Poincar\'e sections, are simply performed when $z$ reaches $2\pi R_{{\rm per}}\mathbb{N}$. 
When the amplitude of the modes, $\epsilon$ and $\epsilon c$ respectively are large enough, 
the two resonances created  by two modes overlap,
and Hamiltonian chaos emerges in the magnetic field line\cite{chirikov1979}.   
Although, even if  there exist two modes, no chaotic field lines  emerged 
when $\epsilon$ is small, (see the case $\epsilon = 0.0005$ in Fig. \ref{fig:mag_c2e-2}).
We can notice that in some situations, we have a region with regular tori 
that cross the entire section, meaning that we have a range of $r$
which acts as an ITB, if the particle trajectories where just bound to follow field lines. 
One question then arises: do these magnetic ITBs works as ITBs for particles?
In order to answer it,
we compute as well Poincar\'e sections of the full particle trajectories. 
It is in this setting that we resort to the ``averaging'' method to compute the sections. 
Indeed we do not have any constant besides the energy that is left, 
moreover this section allows us to have a visual comparison with what happens for the field lines.

\begin{figure}[tb]
	\centering{}
	\begin{tabular}{cc}
	\begin{minipage}{0.45\hsize}
        	\begin{center}         
	{(a)}\\
	\includegraphics[width=7cm]{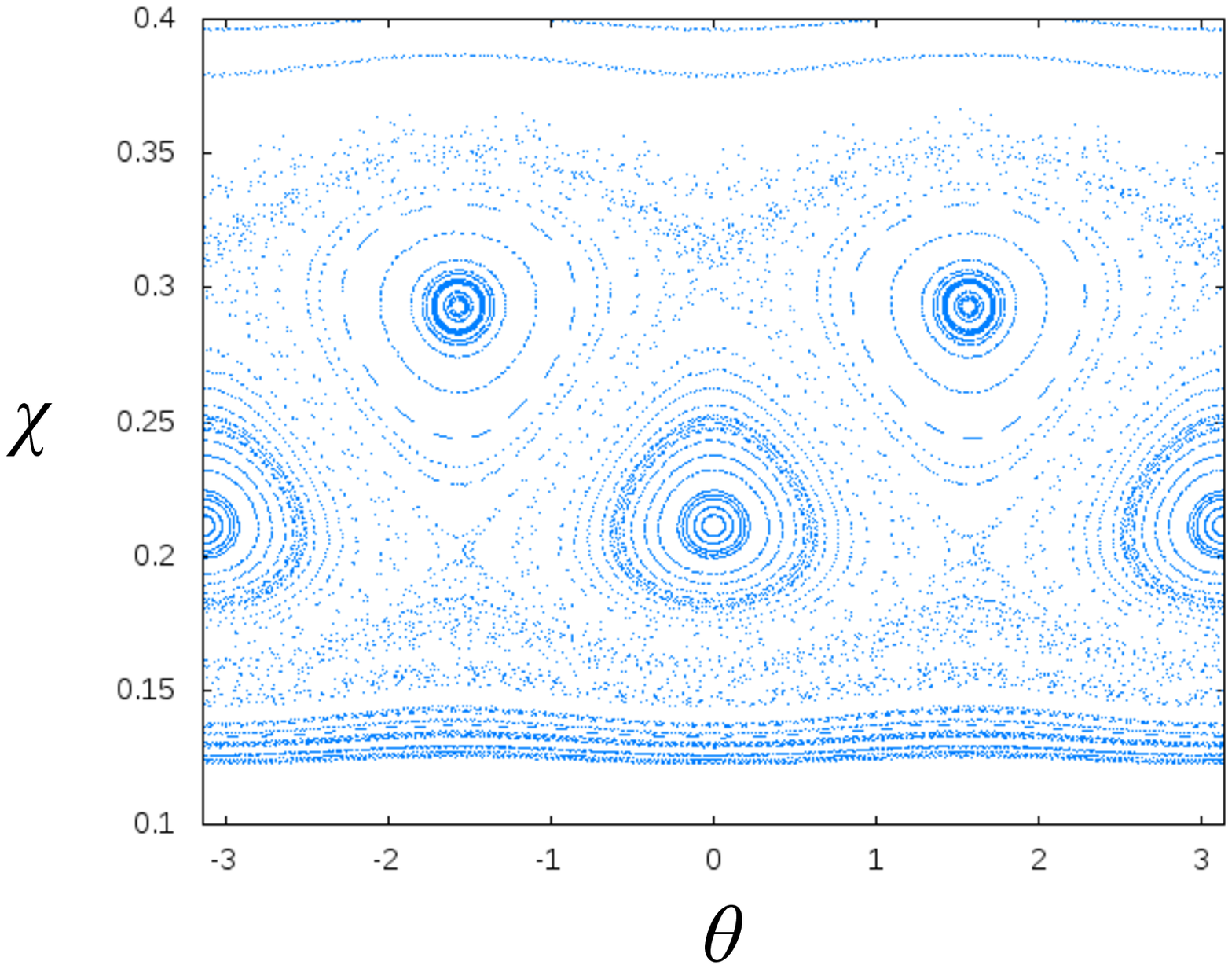}
	\end{center}
      	\end{minipage}	
	\begin{minipage}{0.45\hsize}
        	\begin{center}         
	{(b)}\\	
	\includegraphics[width=7cm]{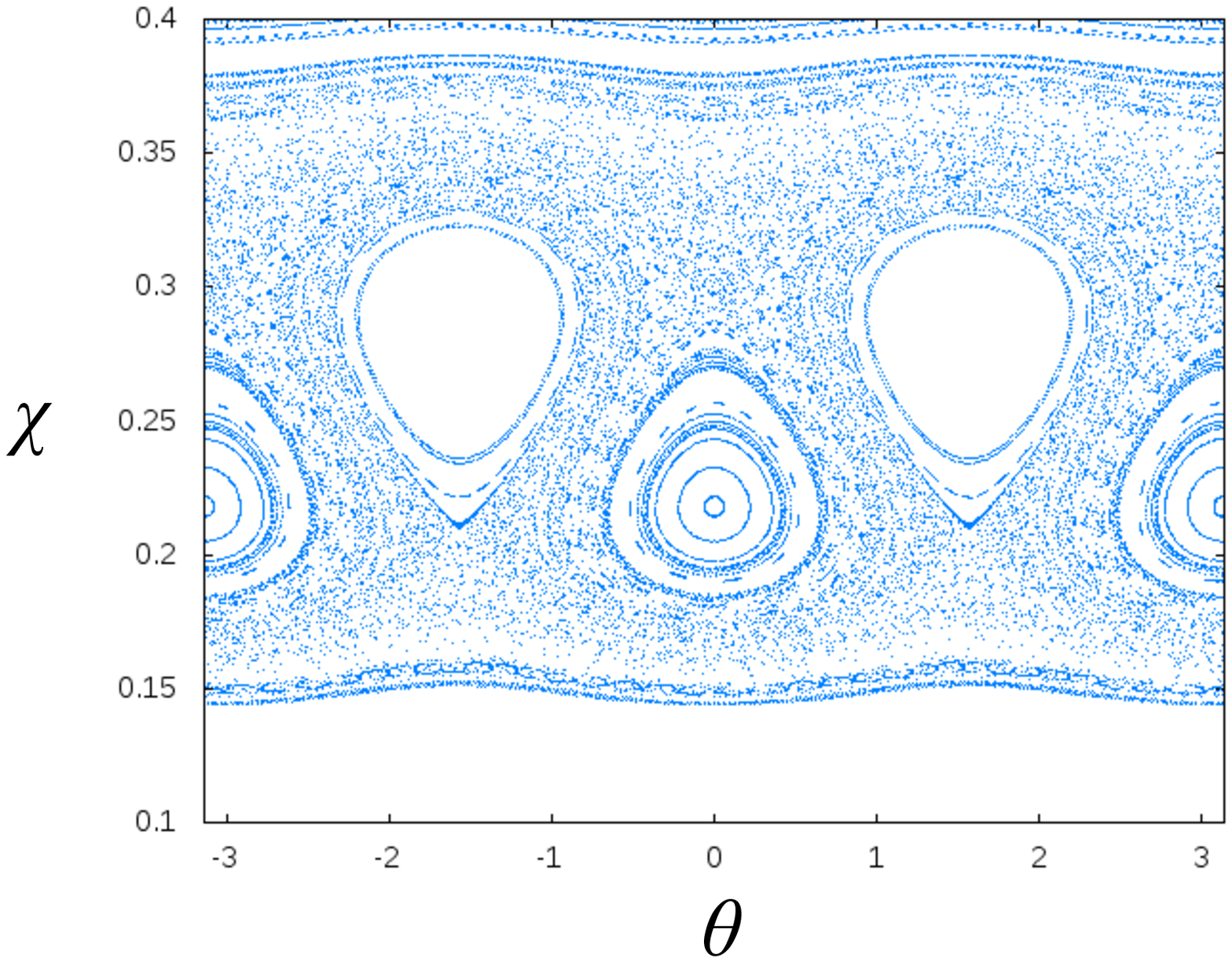}
	\end{center}
      	\end{minipage}\\
	\begin{minipage}{0.45\hsize}
        	\begin{center}         
	{(c)}\\
	\includegraphics[width=7cm]{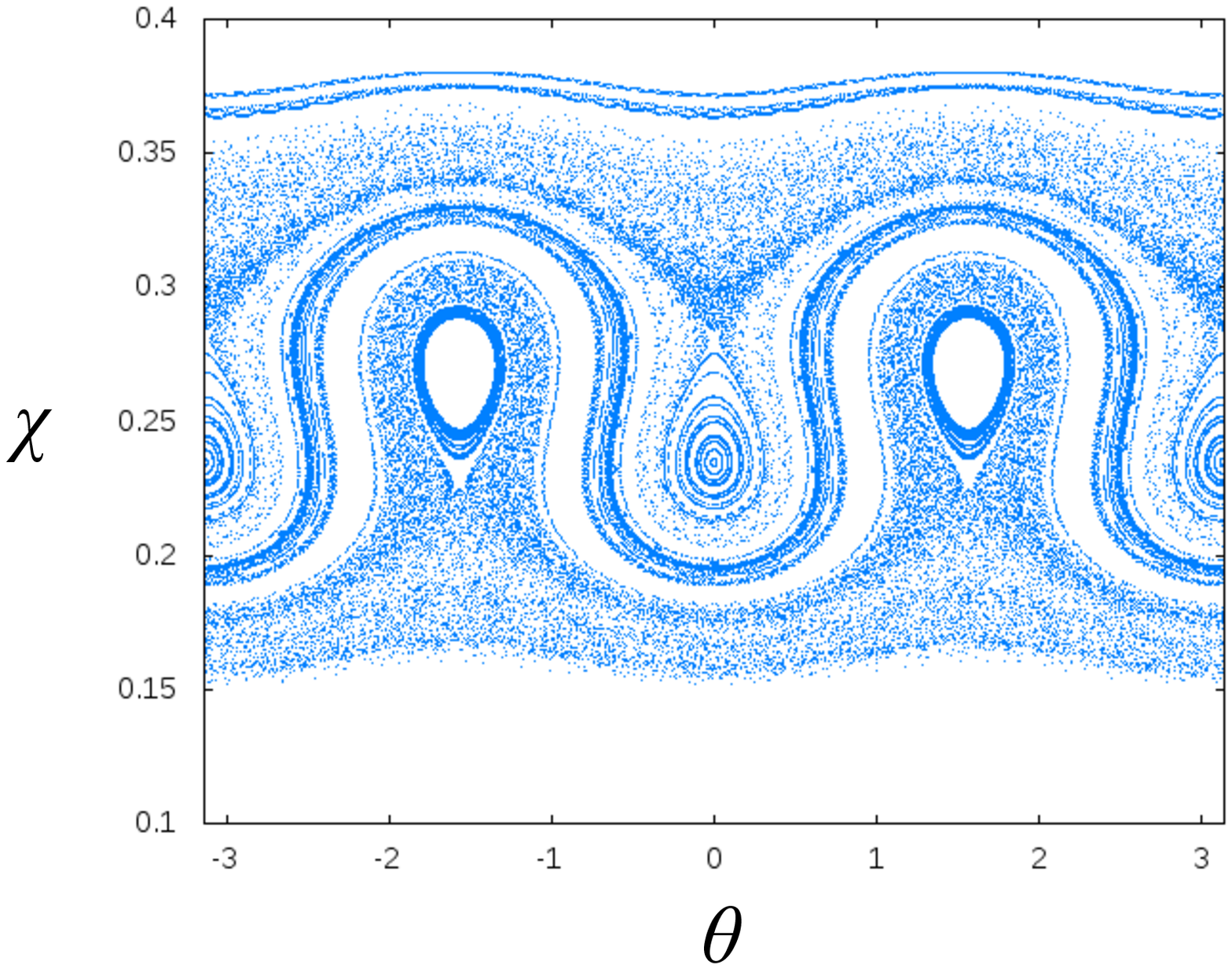}
	\end{center}
      	\end{minipage}
	\begin{minipage}{0.45\hsize}
        	\begin{center}         
	{(d)}\\
	\includegraphics[width=7cm]{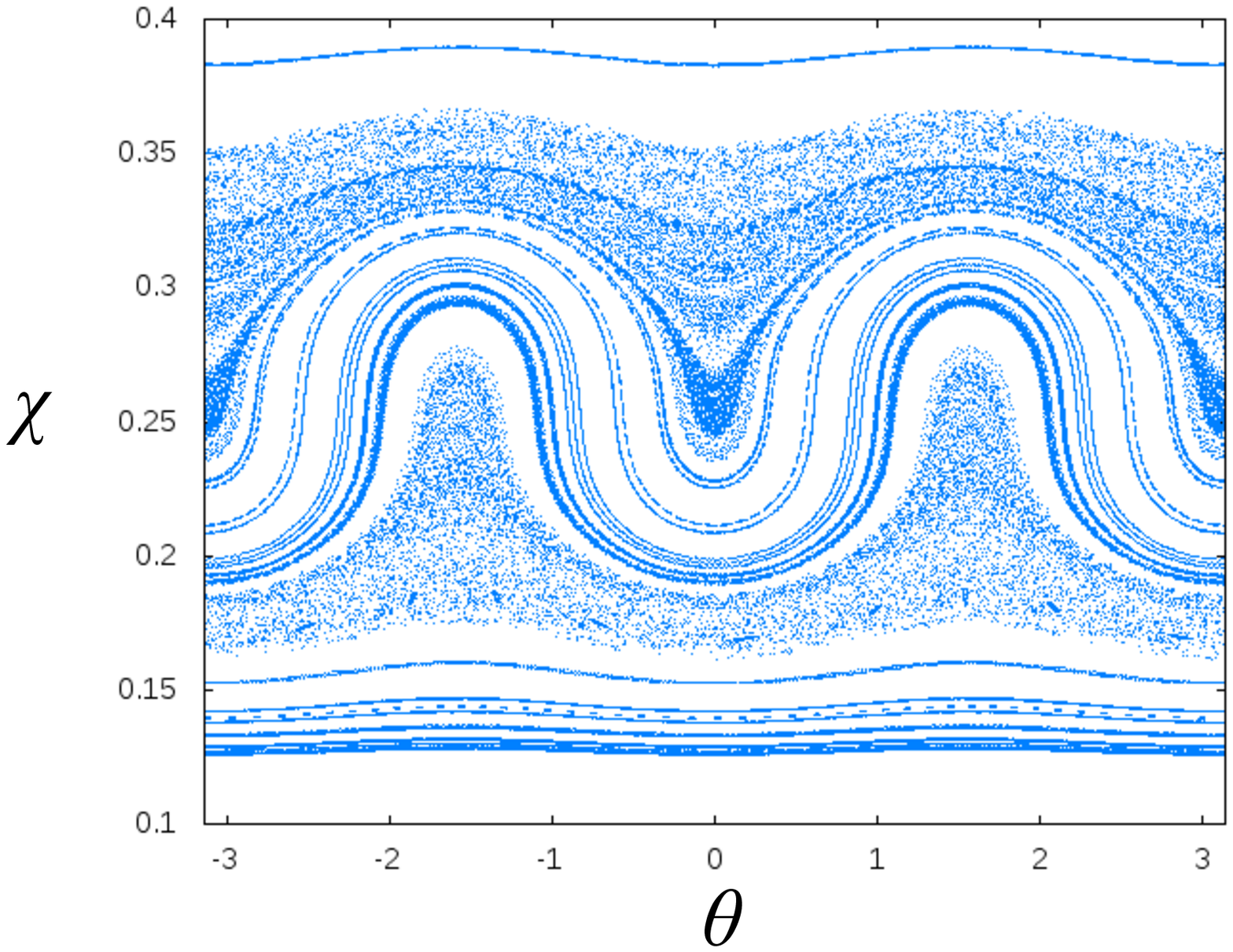}
	\end{center}
      	\end{minipage}
  \end{tabular}
	\caption{Poincar\'e plots obtained from the full orbit with initial pitch angle
	$\phi_{0}=0$, and energy $0.00001$ $0.0001$, $0.0005$ and $0.001$ 
	(respectively 1, 10, 50, and 100keV for proton or alpha particle). 
	We set $\epsilon = 0.0015$ and $c = 0.02$.}
	 \label{fig:particle-tragec-0}
\end{figure}

\begin{figure}[tb]
	\centering{}
	\begin{tabular}{cc}
	\begin{minipage}{0.45\hsize}
        	\begin{center}         
	{(a)}\\
	\includegraphics[width=7cm]{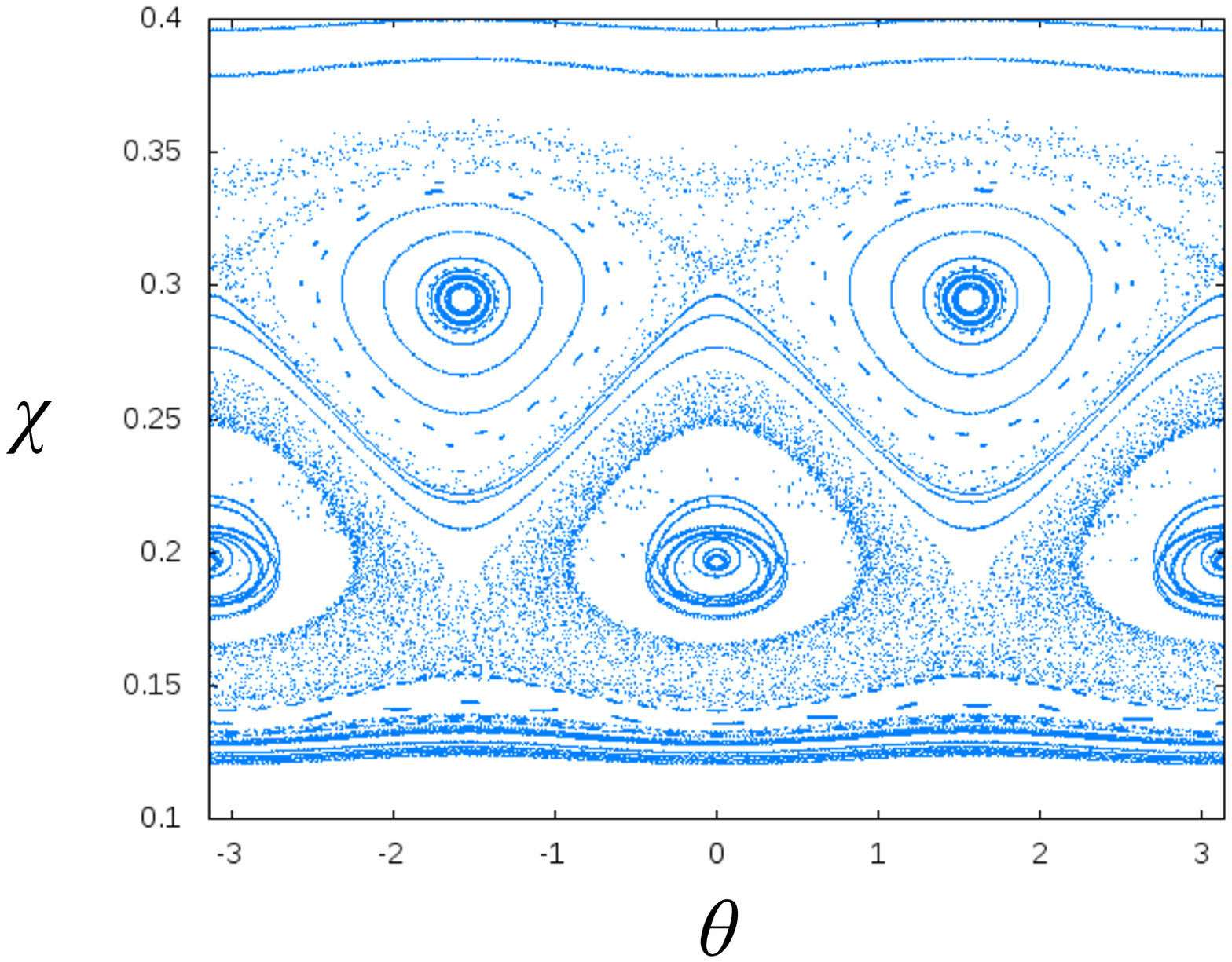}
	\end{center}
      	\end{minipage}
	\begin{minipage}{0.45\hsize}
        	\begin{center}         
	{(b)}\\
	\includegraphics[width=7cm]{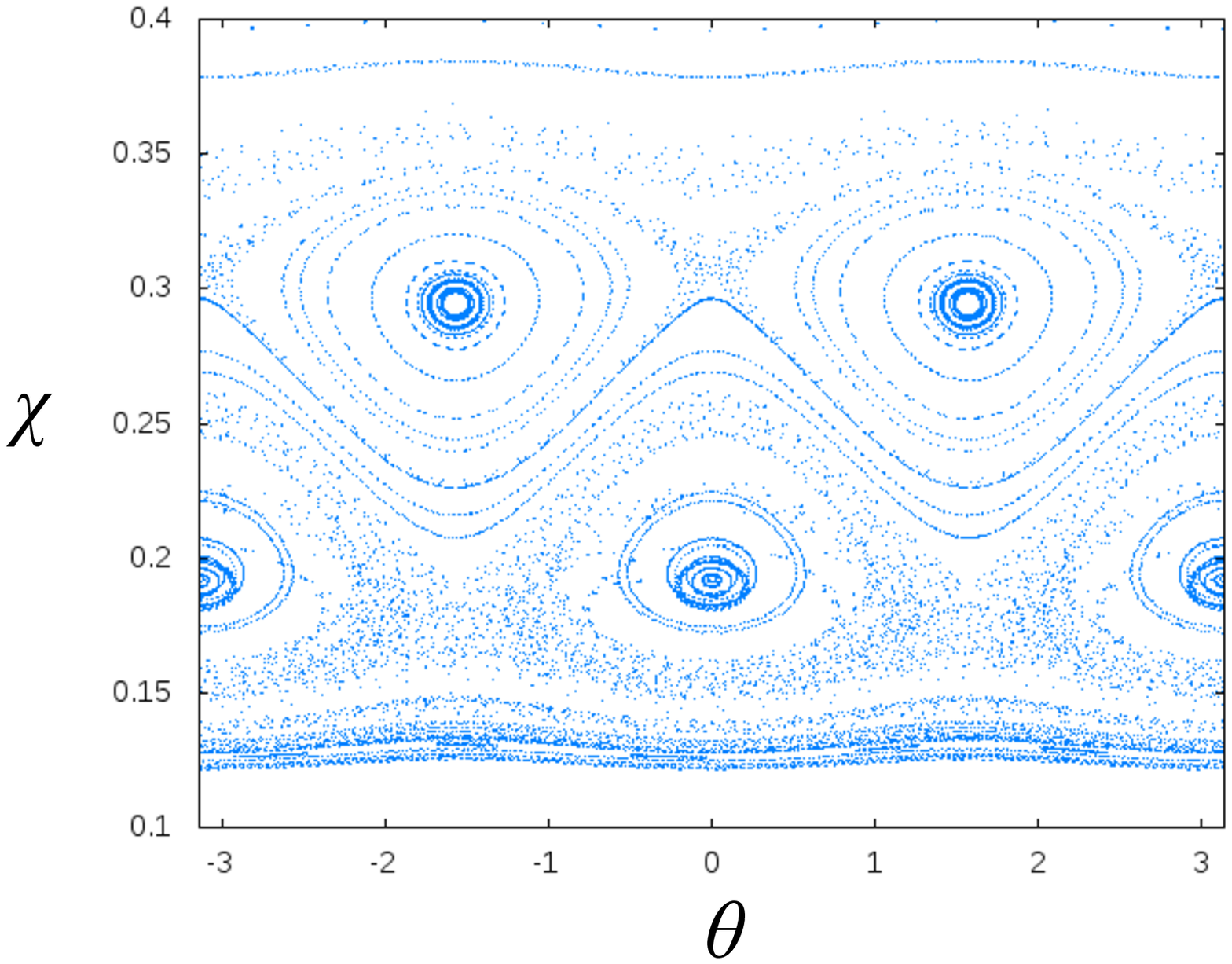}
	\end{center}
      	\end{minipage}
	\\
	\begin{minipage}{0.45\hsize}
        	\begin{center}         
	{(c)}\\
	\includegraphics[width=7cm]{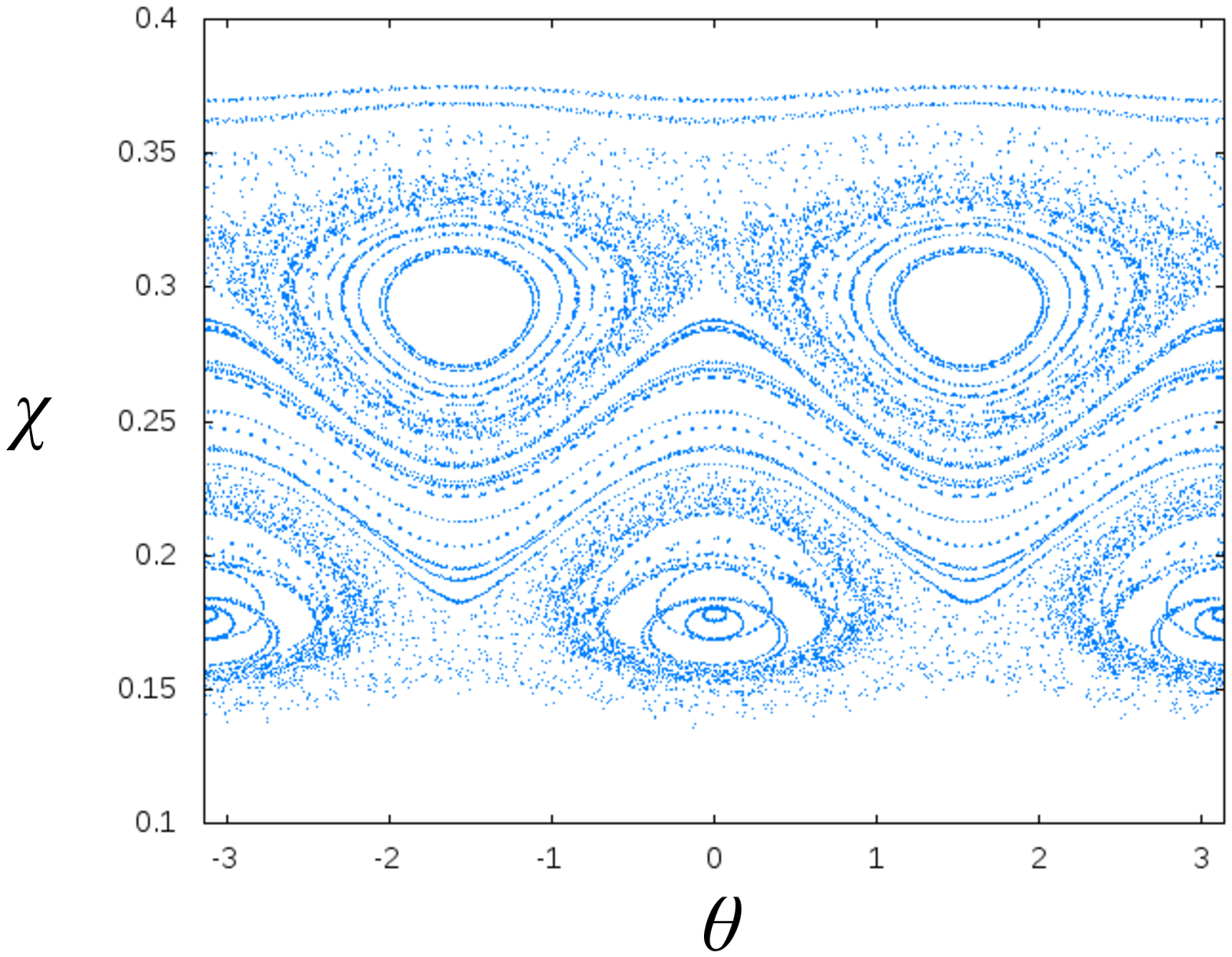}
	\end{center}
      	\end{minipage}
	\begin{minipage}{0.45\hsize}
        	\begin{center}         
	{(d)}\\
	\includegraphics[width=7cm]{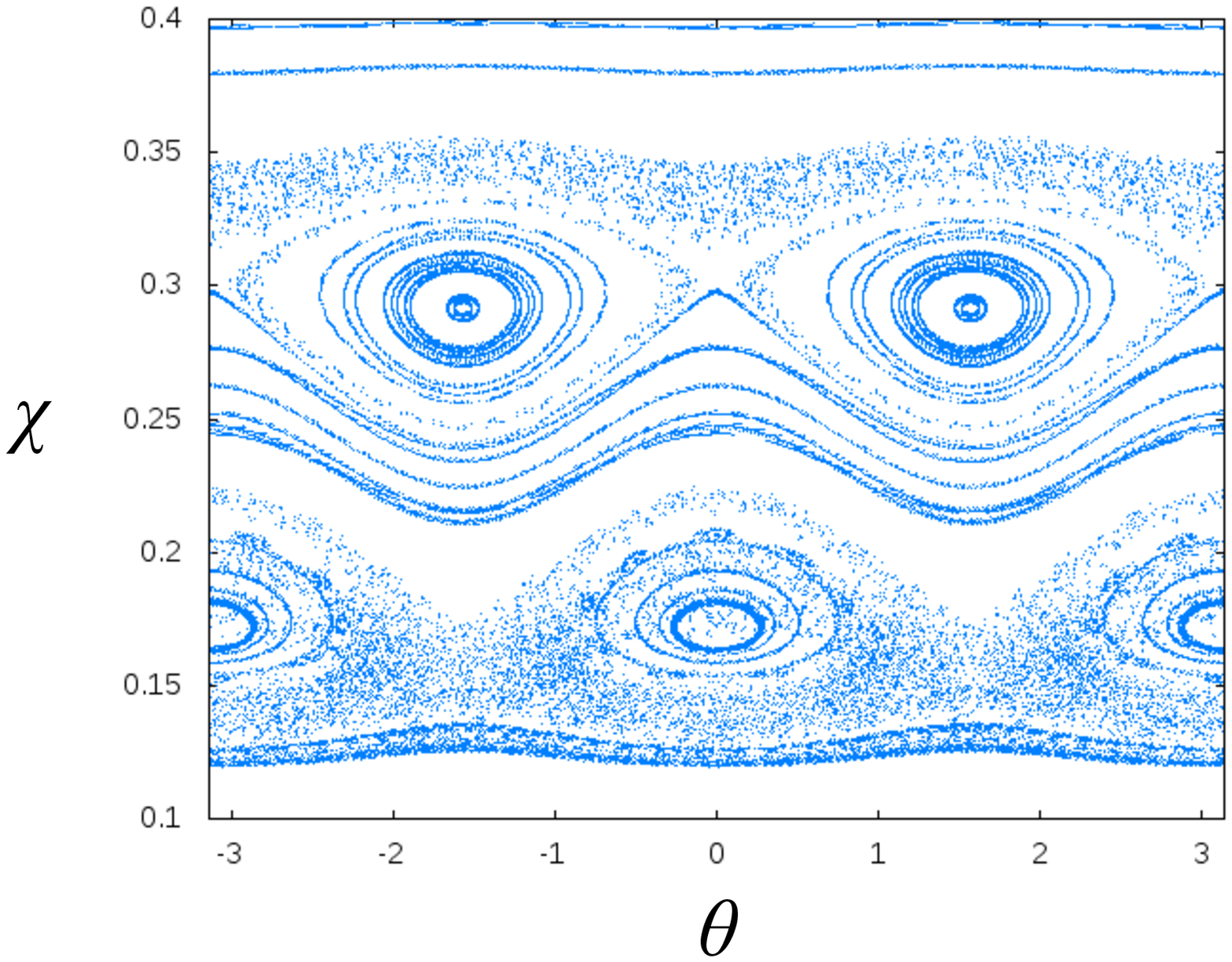}
	\end{center}
      	\end{minipage}
	\\
	\begin{minipage}{0.45\hsize}
        	\begin{center}         
	{(e)}\\
	\includegraphics[width=7cm]{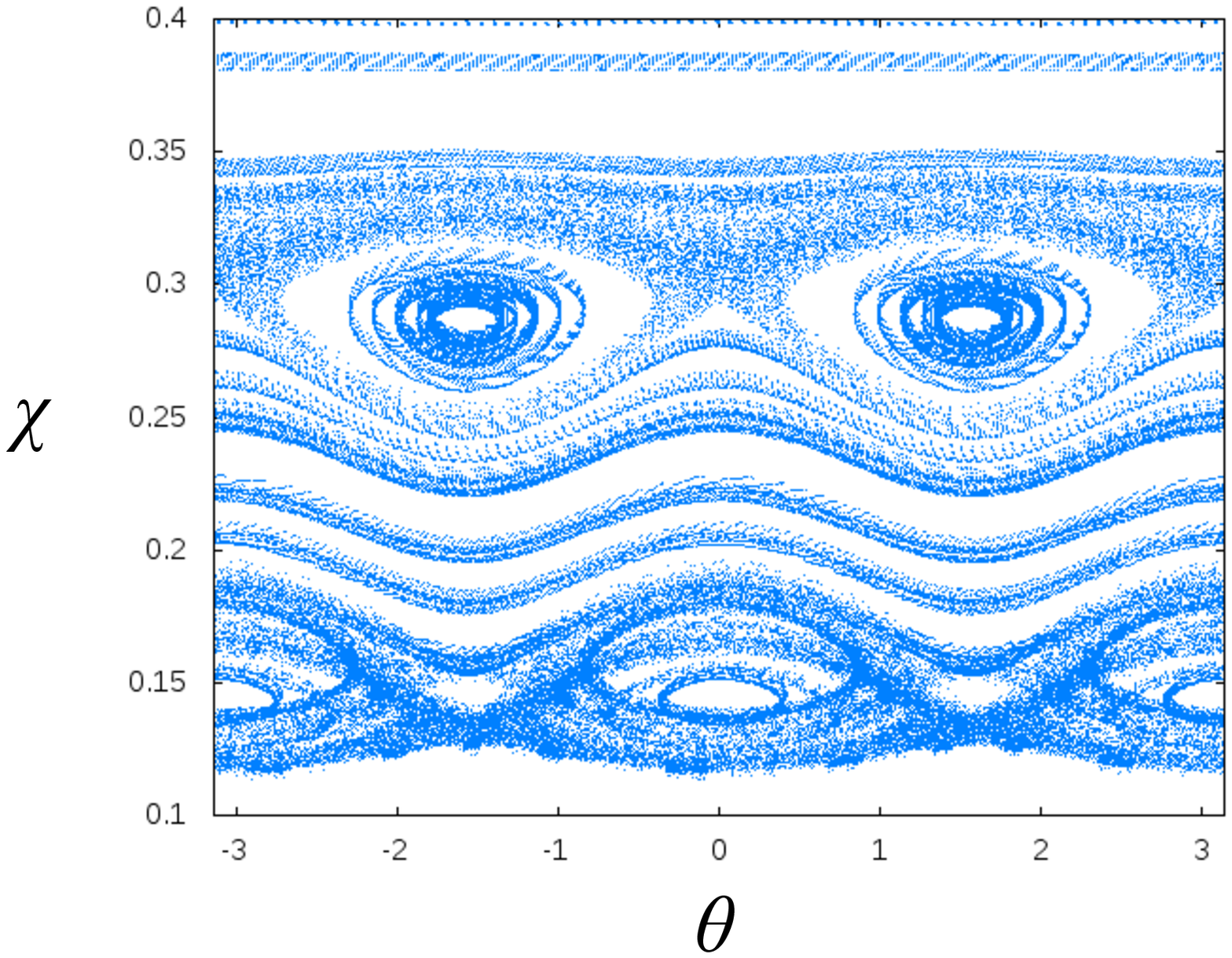}
	\end{center}
      	\end{minipage}
 \end{tabular}

	\caption{Poincar\'e plots obtained from the full orbit with initial pitch angle
	$\phi_{0}=1.25$, and energy 
	0.00005, 0.0001, 0.0005, 0.001, and 0.005. 
	(respectively 1, 10, 50, 100, and 500keV for proton or alpha particle). 
	We set $\epsilon = 0.0015$ and $c=0.02$.}
	\label{fig:particle-tragec-125e-2}
\end{figure}

Before moving on we would like to mention that
there exist several criteria or indices of chaos (stochasticity), 
for example, Melnikov function, Lyapunov characteristic exponent, 
Kolmogorov-Sinai entropy, or Chirikov criterion for resonance overlapping
\cite{Lichtenberg, chirikov1979, Eckmann1985}. 
The chaos of field lines  is explained by the resonance overlapping, but we
only reveal the chaotic nature of trajectories using the tool developed by Poincar\'e, namely Poincar\'e sections
 and do not examine any of the aforementioned criteria for the particle trajectory.
We just look visualized results exhibited in 
Figs.~\ref{fig:particle-tragec-0} and \ref{fig:particle-tragec-125e-2}
and discuss about 
qualitative difference between the field lines and the particle trajectories. 
Regarding chaos per se, as the presence of a chaotic sea
does not prevent the system from eventually having a zero Lyapunov exponent and displaying so called weak chaos
features\cite{Gaspard1988,Korabei2009}. They are though sufficient to show that the system is not integrable.

The initial condition of the particles are defined by 
the kinetic energy $E=\vec{v}^{2}/2$, initial position $\mathbf{q}_{0}$,
and the pitch angle $\phi_{0}$ between initial velocity $\mathbf{v}_{0}$
and the magnetic field at the initial point $\mathbf{B}(\mathbf{q}_{0})$:
\begin{equation}
	\begin{split}
		\mathbf{v}_{\parallel}=\sqrt{2E}\cos\phi_{0}\frac{\mathbf{B}}{\|\mathbf{B}\|},\quad\mathbf{v}_{\perp} 
		& =\sqrt{2E}\sin\phi_{0}\frac{\mathbf{e}_{z}\wedge\mathbf{B}}{\|\mathbf{e}_{z}\wedge\mathbf{B}\|}. 
	\end{split}
\end{equation}
The particle trajectories are exhibited in Figs.~\ref{fig:particle-tragec-0}
and \ref{fig:particle-tragec-125e-2} for different values of energy
$E$ and different pitch angle $\phi_{0}$.

Thick regular regions dividing the chaotic region appear in two cases.
The first case is when the energy $E$ is large and the initial pitch angle $\phi_{0}$ is small. 
In this case, we cannot find an island around the elliptical critical point which is found in 
the Poincar\'e plot of the magnetic field line (Fig.~\ref{fig:mag_c2e-2}). 
Further, the topological structure of the particle trajectory strongly depends on the energy. 
The other case is when the initial pitch angle $\phi_{0}$ is large. 
There is then a thick regular region between two stochastic regions including two islands. 
Unlike the small pitch angle case, the topological structure of the trajectory does not depend on the energy. 
In both cases, the shape of the particle trajectory is completely different from the magnetic field line profile.

Let us investigate what happens in the small initial pitch angle case.
When the kinetic energy is quite low, the particle moves along the
magnetic field and the topological structure of the trajectory is similar to the field line profile. 
When the energy increases, we notice that the particle moves as if it effectively feels a magnetic field
with larger $\epsilon$ (see Figs.~\ref{fig:mag_c2e-2} and \ref{fig:mag_c0}).
Let $\mathbf{v}_{\parallel}^{0}$ and $\mathbf{v}_{\perp}^{0}$ be velocity parallel and perpendicular
to $\mathbf{b}^{0}=\mathbf{B}_{0}/\|\mathbf{B}_{0}\|$ (the unit vector
along the magnetic line). The particle motion is determined by the
equation ${\rm d}\mathbf{v}/{\rm d}t=\mathbf{v}\wedge\mathbf{B}$. Due to the
small pitch angle, we can assume $\|\mathbf{v}_{\perp}^{0}\|\ll\|\mathbf{v}_{\parallel}^{0}\|$, 
and conclude
\begin{equation}
	\|\mathbf{v}_{\perp}^{0}\wedge\mathbf{B}_{0}\|
	\gtrsim\epsilon\|\mathbf{v}_{\parallel}^{0}\wedge\mathbf{B}_{1}\|
	\gg\epsilon\|\mathbf{v}_{\perp}^{0}\wedge\mathbf{B}_{1}\|,\label{eq:case1}
\end{equation} 
where $\mathbf{v}_{\parallel}^{0} \wedge \mathbf{B}_0 = 0$ by definition. 
The leading term
of the Lorentz force is hence
\begin{equation}
	\mathbf{v}\wedge\mathbf{B}\simeq
	\mathbf{v}_{\perp}^{0}\wedge\mathbf{B}_{0}+\epsilon\mathbf{v}_{\parallel}^{0}\wedge\mathbf{B}_{1}.
\end{equation}
The first term of the right-hand-side gives the unperturbed motion,
and the second one the perturbation effect. 
If the parallel velocity $\|\mathbf{v}_{\parallel}^{0}\|$ is larger, 
the perturbation term 
$\epsilon\mathbf{v}_{\parallel}^{0}\wedge\mathbf{B}_{1}$ is relatively larger compered with the unperturbed part.
In other words, the weight of the perturbation in the equation of motion becomes heavier than the unperturbed part.
The particle therefore effectively feels a magnetic field with a large $\epsilon$ 
as shown in Fig.~\ref{fig:particle-tragec-0}. 
The change in the particle trajectory as the energy increases is similar to 
the change of the magnetic field trajectories, (heteroclinic like $\to$ homoclinic like)
as $\epsilon$ increasing in the magnetic field lines (see. Fig.~\ref{fig:mag_c0}).
Indeed, for low energy ($E=0.00001$), the particle trajectories are very similar to the field line trajectories. 
We can see the homoclinic like structure when
the perturbation $\epsilon$ is large in Figs.~\ref{fig:mag_c2e-2} and \ref{fig:mag_c0}, 
and the plot for $E=0.0001$ in Fig.~\ref{fig:particle-tragec-0} is similar to them. 
As the energy increases, 
this is not the case, because the drift effects become dominant 
\cite{Boozer2004,Jackson98}. 
In particular, the leading order term of the curvature drift velocity, 
\begin{equation}
	\mathbf{v}_{\rm cd}^0 = \frac{\|\mathbf{v}_{\parallel}^{0}\|^2}{\omega_{\rm gyr}^0}\mathbf{b}_0 \wedge \mathbf{\kappa}_0, \quad
	\mathbf{\kappa} = \mathbf{b}_0 \cdot \nabla \mathbf{b}_0, 
\end{equation}
is proportional to $\|\mathbf{v}_{\parallel}^{0}\|^2$, 
where $\omega_{\rm gyr}$ denotes the gyro-frequency. 
As shown in Fig.~\ref{fig:particle-tragec-0} for $E=0.0005$, 
a regular region that acts as an ITB appears between two chaotic regions. 
As the energy increases, perturbation effect is suppressed and the ITB becomes wider.

When the pitch angle is large, 
$\|\mathbf{v}_{\perp}^{0}\|>\|\mathbf{v}_{\parallel}\|$
and 
\begin{equation}
	\|\mathbf{v}_{\perp}^{0}\wedge\mathbf{B}_{0}\|
	\gg\epsilon\|\mathbf{v}_{\perp}^{0}\wedge\mathbf{B}_{1}\|
	\gtrsim\epsilon\|\mathbf{v}_{\parallel}^{0}\wedge\mathbf{B}_{1}\|. \label{eq:case2}
\end{equation}
Unlike in case in Eq. \eqref{eq:case1},
the ratio between the unperturbed part and perturbation part in the Lorentz force 
is independent of the energy. 
Therefore, the energy cannot affect the topology of
the particle trajectories as strongly as in the low energy case in Eq. \eqref{eq:case1}. 
However, as the energy increases, the Larmor radius becomes large, 
the particle gyrates quickly, and the perturbation is averaged and wiped out. 
In this case, the structure of the particle orbits 
quantitatively changes as the energy increases, 
the chaotic region gets to be smaller
and the ITB appears and becomes wider.  

It should be remarked that it has already been discussed that finite Larmor radius effects
suppress the chaotic behavior in different situation \cite{Castillo2012,Martinell2013}. 
In this paper we have
dealt with the six dimensional system with a magnetic field having
several modes and a trivial electric field $\mathbf{E}=0$. 
On the other hand, in del-Castillo-Negrete and Martinell\cite{Castillo2012,Martinell2013}, 
the similar perturbation suppression by the finite Larmor radius effect has been
considered for a three dimensional system; an E $\times$ B test particle model, 
with an ideal magnetic field and the perturbation was introduced
through some modes of the electric field.

\begin{table}[tb]
    \centering
	\renewcommand{\arraystretch}{1.25}
	\begin{tabular}{r||c|c}
		\hline
		\hline
			& Small initial pitch angle & Large initial pitch angle  \\
		\hline
		\hline
		\shortstack{High\\energy}& 
		\multicolumn{2}{|c}{
			\shortstack{
			Effect of magnetic field perturbation is suppressed, 
			}
		}
		\\ \cline{1-2}
		\shortstack{$\qquad$\\ {\huge $\Uparrow$}\\$\qquad$}& 
		\shortstack{
			Effect of magnetic field \\ perturbation 
			in trajectory \\
			becomes larger.
		}
		&
		\shortstack{and \\
			ITB appears and \\ gets to be wider.
		}
		\\
		\hline
		\shortstack{Low\\energy}&  
		\multicolumn{2}{|c}{Particles move along magnetic field line.}
		\\
		\hline
		\hline
	\end{tabular}
	     \label{tab:sum}
	\caption{
    		Summary of topology change
    	}
\end{table}

\section{Conclusion and Remark}

We have studied charged particle motion in a magnetic field in cylindric geometry. 
Electric field and radiation of the particle 
and back reaction effects have been neglected and will be studied in a future publication. 

In the first part, we have shown the existence of chaotic particle motion in regular integrable magnetic fields. 
To study this phenomenon, we have constructed Poincar\'e sections on the iso-$p_{\theta}$ plane. 
We have derived the effective Hamiltonian $H_{{\rm eff}}$ from the unperturbed Hamiltonian
directly from the full Hamiltonian unperturbed magnetic field depending only on $r$.
We have then considered adding a magnetic perturbation with only one mode. 
In this setting, the original two invariants of the unperturbed
system, $p_{\theta}$ and $p_{z}$, oscillate and so does the separatrix
of the effective Hamiltonian $H_{{\rm eff}}$. 
As a consequence of separatrix crossing, 
Hamiltonian chaos arises with the formation of a stochastic layer \cite{Tennyson86,Cary1986,Neishtadt86}. 
We have shown that the particles close to the separatrix of $H_{{\rm eff}}$ 
in the $(r,p_{r})$ plane display stochastic behavior, 
and the width of the stochastic zone scales like $\sqrt{\epsilon}$ as expected.
Particles far from this region move regularly,
This result allows to estimate the density of particles having chaotic trajectories in the integrable field, 
which might be useful for estimating potential errors introduced by global gyrokinetic reductions. 

In the second part, we have discussed the link between the topology
of the magnetic field lines and the topology of the particle trajectories.
In this case, it is necessary to take a Poincar\'e plot on the plane $z\in2\pi\mathbb{Z}$ 
to compare the particle motion with the field line orbits. 
If one takes Poincar\'e plots naively, it is hard to observe
what actually happens, due to the finite Larmor radius. 
To address this problem, 
we have proposed\ a method to construct
effective ``Poincar\'e plots''. 
Using this method, we have found that topology of the magnetic field lines 
and the topology of particle trajectories in the real space can be 
significantly different. 
When the particle energy is high and the pitch angle is small,
the structure of the particle trajectory is similar to the one of the magnetic field line with an
effective perturbation $\epsilon_{{\rm eff}}$ larger than the actual one $\epsilon$. 
This phenomenon has been explained by the balance of 
the unperturbed term $\mathbf{v}_{\perp}^{0}\wedge\mathbf{B}_{0}$
and the dominant perturbation term $\epsilon\mathbf{v}_{\parallel}^{0}\wedge\mathbf{B}_{1}$.
A small pitch angle implies that the energy scales as $H\simeq\|\mathbf{v}_{\parallel}^{0}\|^{2}/2$,
and the perturbation term gets to be larger as $H$ becomes larger.
However, when the pitch angle is large,
the shape of particle trajectories seems to be independent on the energy.
This is because the ratio of the unperturbed part and the perturbed
part in the Lorentz force is not so drastically changed when the energy
increases unlike the first case. 
Qualitative investigations on the large finite Larmor radius effect appear to be necessary. 
We showed that even though magnetic barrier is destroyed, 
the ITB actually remained effective for a given energy range of the particles. 
On the one hand, it is possible that particles 
with small pitch angle and with some values of kinetic energy destroy the magnetic ITB, 
because the structure of the particle orbit is changed qualitatively 
as we have exhibited in Fig. \ref{fig:particle-tragec-0}. 
This does not happen for particles with large pitch angle. 
In this case, the ITB acts more like a filter than a real barrier.
More precise studies, like taking into account the effects of the electric field are important
in order to consider this result relevant for the confinement of fusion plasmas,
but these preliminary results are promising.
\acknowledgements
This work has been carried out thanks to the support of the A$\ast$MIDEX
project (n$^\circ$ ANR-11-IDEX-0001-02) funded by the ``investissements d'Avenir''
French Government program, managed by the French National Research Agency
(ANR).
DdcN acknowledges support from the Office of Fusion Energy Sciences 
of the US Department of 
Energy at Oak Ridge National Laboratory, managed by UT-Battelle, LLC, 
for the U.S.Department of Energy under contract DE-AC05-00OR22725.



\begin{thebibliography}{10}
\bibitem{Boozer2004} A. H. Boozer, Rev. Mod. Phys. \textbf{76}, 1071
(2004).

\bibitem{CaryBrizard2009} J. R. Cary and A. J. Brizard, Rev. Mod.
Phys. \textbf{81}, 693 (2009).

\bibitem{Littlejohn1981}
R. G. Littlejohn, Phys. Fuids, {\bf 24}, 1730 (1981). 

\bibitem{Lichtenberg}
 A. J. Lichtenberg and M. A. Lieberman, 
Regular and Chaotic Dynamics, {\it 2nd edition}
(Springer-Verlag, New York, 1992). 


\bibitem{Landsman2004}
A. S. Landsman, S. A. Cohen, and A. H. Glasser, 
Phys. Plasmas {\bf 11}, 947 (2004).

\bibitem{Cambon2014} B. Cambon, X. Leoncini, M. Vittot, R. Dumont,
and X. Garbet, Chaos \textbf{24}, 033101 (2014).

\bibitem{Pfefferle2015}
D. Pfefferl\'e, J. P. Graves, and W. A. Cooper,
Plasma Phys. Control. Fusion {\bf 57}, 054017 (2015).

\bibitem{Constantinescu2012}
D. Constantinescu and M.-C. Firpo, 
Nucl. Fusion {\bf 52}, 054006 (2012).

\bibitem{Connor2004}
J. W. Connor, T. Fukuda, X. Garbet, C. Gormezano,
V. Mukhovatov, M. Wakatani, 
the ITB Database Group and the ITPA Topical Group on Transport and Internal Barrier Physics,
Nucl. Fusion {\bf 44}, R1 (2004). 

\bibitem{delcastillo96} D. del-Castillo-Negrete, J.M. Greene, and P.J. Morrison, Physica D {\bf 91} 1 (1996).

\bibitem{delcastillo2000} D. del-Castillo-Negrete: Phys. of Plasmas {\bf 7}, 1702 (2000).

\bibitem{Balescu98} R. Balescu,  Phys. Rev. E, {\bf 58}, 3781 (1998).


\bibitem{delcastillo97}
D. del-Castillo-Negrete, J. M. Greene, and P. J. Morrison, 
Physica D, {\bf 100}, 311 (1997).


\bibitem{Morrison2009}
P. J. Morrison and A. Wurm, 
Nontwist maps, Scholarpedia, {\bf 4}, 3551 (2009).

\bibitem{delsham2000}
A. Delsham and R. de la Llave, SIAM J. Math. Anal., {\bf 31}, 1235 (2000).

\bibitem{gonzalez2014}
A. Gonz\'alez-Enr\'iquez, A. Haro, and R. de la Llave, 
Singularity Theory for Non-Twist KAM Tori, 
Memoirs of the American mathematical society, {\bf 227} (2014).

\bibitem{Cary83} J. R. Cary and R. G. Littlejohn, Ann. Phys. \textbf{151},1 (1983).
\bibitem{Abdullaev2013} 
S. S. Abdullaev, 
Magnetic Stochasticity in Magnetically Confined Fusion Plasmas: 
Chaos Field Lines and Charged Particle Dynamics, 
(Springer, 2013).   


\bibitem{Landau-mechanics}
L. D. Landau,and E. M. Lifshitz, 
Mechanics, {\it 2nd edition}
(Pergamon Press, Bristol, 1969).


\bibitem{Blazevski2013} D. Blazevski and D. del-Castillo-Negrete,
Phys .Rev. E \textbf{87}, 063106 (2013).

\bibitem{MacLchlan92} R. I. McLachlan and P. Atela, Nonlinearity
\textbf{5}, 541 (1992).

\bibitem{Neishtadt86} A. I. Neishtadt, Prikl. Metm, Mekhan. USSR,
\textbf{51}, 586 (1987).

\bibitem{Tennyson86} J. L. Tennyson, J. R. Cary, and D. F. Escande,
Phys. Rev. Lett. \textbf{56}, 2117 (1986).

\bibitem{Cary1986} J. R. Cary, D. F. Escande, and J. L. Tennyson
Phys. Rev. A \textbf{34}, 4256 (1986).

\bibitem{Leoncini2009} X. Leoncini, A. Neishtadt, and A. Vasiliev,
Phys. Rev. E \textbf{79}, 026213 (2009).

\bibitem{chirikov1979}
B. V. Chirikov, Phys. Rep. {\bf 52}, 263 (1979).

\bibitem{Eckmann1985}
J. P. Eckmann and D. Ruelle, Rev. Mod. Phys. {\bf 57}, 617 (1985).

\bibitem{Gaspard1988}
P. Gaspard and X.-J Wang, Proc. Natl. Acad. Sci. {\bf 85}, 4591 (1988). 

\bibitem{Korabei2009}
N. Korabel and E. Barkai,
Phys. Rev. Lett. {\bf 102}, 050601, (2009).

\bibitem{Jackson98} J. D. Jackson, Classical Electrodynamics, \textit{3rd edition} (Wiley, USA, 1998).

\bibitem{Castillo2012} D. del-Castillo-Negrete and J. J. Martinell,
Commun. Nonlinear Sci, Numer. Simulat. \textbf{17}, 2031 (2012).

\bibitem{Martinell2013} J. J. Martinell and D. del-Castillo-Negretee,
Phys. Plasmas \textbf{20}, 022303 (2013).
\end{thebibliography}
\end{document}